\documentclass[10 pt, letterpaper]{article}

\usepackage{amssymb}
\usepackage{amsmath}
\usepackage{graphicx}
\usepackage{times}
\usepackage[T1]{fontenc}

\newcommand\highertop{\rule{0pt}{3.1ex}}

\begin{document}
\date{}

\title{On sensor fusion for airborne wind energy systems
\thanks{This manuscript is a preprint of a paper submitted for possible publication on the IEEE Transactions on Control Systems Technology and is subject to IEEE Copyright. If accepted, the copy of
record will be available at \textbf{IEEEXplore} library: http://ieeexplore.ieee.org/.}
\thanks{This research has received funding from the California Energy Commission under the EISG grant n. 56983A/10-15 ``Autonomous flexible wings for high-altitude wind energy generation'', and from the European Union Seventh Framework Programme (FP7/2007-2013) under grant agreement n. PIOF-GA-2009-252284 - Marie Curie project ``Innovative Control, Identification and Estimation Methodologies for Sustainable Energy Technologies''. The authors acknowledge SpeedGoat$^\circledR$'s Greengoat program.}}
\author{L. Fagiano\thanks{Corresponding author: fagiano@control.ee.ethz.ch.}, K. Huynh, B. Bamieh, and M. Khammash
\thanks{The authors are with the Dept. of Mechanical Engineering, University of California at Santa Barbara, CA, USA. L. Fagiano is also with the Automatic Control Laboratory, Swiss Federal Institute of Technology, Zurich, Switzerland. M. Khammash is also with the Department of Biosystems Science and Engineering, Swiss Federal Institute of Technology, Zurich, Switzerland.}}
\maketitle

\begin{abstract}
A study on filtering aspects of airborne wind energy generators is presented. This class of renewable energy systems aims to convert the aerodynamic forces generated by tethered wings, flying in closed paths transverse to the wind flow, into electricity.
The accurate reconstruction of the wing's position, velocity and heading is of fundamental importance for the automatic control of these kinds of systems.
The difficulty of the estimation problem arises from the nonlinear dynamics, wide speed range, large accelerations and fast changes of direction that the wing experiences during operation. It is shown that the overall nonlinear system has a specific structure allowing its partitioning into sub-systems, hence leading to a series of simpler filtering problems. Different sensor setups are then considered, and the related sensor fusion algorithms are presented.
The  results of experimental tests carried out with a small-scale prototype and wings of different sizes are discussed. The designed filtering algorithms rely purely on kinematic laws, hence they are independent from features like wing area, aerodynamic efficiency, mass, etc. Therefore, the presented results are representative also of systems with larger size and different wing design, different number of tethers and/or rigid wings.
\end{abstract}

\section{Introduction}\label{S:intro}
High-altitude wind is a vast, still untapped renewable source of energy that has received an increasing attention in the last decade, from both industry and academia, with the ongoing development of a series of technologies that fall under the umbrella name of airborne wind energy. While many concepts of airborne wind energy generators were present already in patents and publications in the late 1970s \cite{Mana76,FlRo79,Loyd80}, it is only in recent years that an increasing number of research groups and companies started to develop operating prototypes to convert the energy of high-altitude wind, blowing up to 1000 m from the ground, into electricity (see e.g. \cite{Makani,skysails,ampyx,enerkite,windlift,kitenergy,CaFM07,IlHD07,CaFM09c,TBSO11,BaOc12,Altaeros12}). The activities carried out worldwide in the last six years allowed to define and assess, through theoretical, numerical and experimental research, some common grounds of airborne wind energy (see \cite{FaMi12} for an overview), as well as its potentials to provide cheap electricity in large quantities, available practically everywhere in the world \cite{FaMP09,phd_thesis_fagiano}.

Airborne wind energy generators exploit the motion of wings flying fast in the so-called crosswind conditions, i.e. roughly perpendicularly to the wind flow, and linked to the ground by flexible lines. The aerodynamic forces generated by the wing are then converted into electricity using one of several possible approaches \cite{FaMi12}. Control engineering plays a crucial role in airborne wind energy technologies, since, differently from conventional wind energy based on wind turbines, there is no passive, rigid structure that constrains the path of the wing and imposes its crosswind motion. Rather, this task has to be accomplished by an active control system that keeps track of the wing trajectory and issues suitable control inputs through actuators. In the last six years, the problem of control design for airborne wind energy generators has been studied by several research groups and companies, leading to a quite significant series of theoretical and numerical studies (see e.g. \cite{IlHD07,WiLO08,CaFM09c,FaMP11,BaOc12,phd_thesis_fagiano}) as well as experimental tests, of which few works in the literature report  measured data \cite{CaFM09c,FaMP09,ErSt12,FZMK13_arxiv}.

All of the mentioned control approaches rely on the availability of a series of variables to be used for feedback, most notably the wing's three-dimensional position and velocity. The problem of estimating with sufficiently good accuracy and limited lag these quantities is therefore of paramount importance in the field, however in the literature there are only few works on this topic, highlighting the specific issues that need to be addressed and providing either numerical or experimental results (see e.g. \cite{ErStECC2013}, concerned with the navigation algorithms of large towing kites for seagoing vessels).

In this paper, we contribute to fill this gap by providing a formulation of the estimation problem and by analyzing two different sensor setups and the related sensor fusion algorithms. In particular, the first sensor setup implies the use of a commercial Inertial Measurement Unit (IMU), installed on the wing and equipped with  accelerometers, gyroscopes, magnetometers, a GPS, and a barometer. We propose two algorithms for this sensor setup: the first approach is a quite standard linear observer, while the second one introduces a nonlinear correction to account for the specific kinematic constraints of the considered application.
In the second sensor setup, we consider the use of a line angle measurement system in addition to the accelerometers, gyroscopes and magnetometers.
The presented study is focused on  the estimation of the wing's position and of its ``velocity angle'', which is a variable well-suited for automatic control strategies, when the wing is flying fast in crosswind conditions (see 
\cite{FZMK13_arxiv} for more details on the automatic control design). We present  experimental results obtained with a small-scale prototype operating with a fixed line length of 30 m and equipped with an IMU and a line angle measurement system, which we developed and built ad-hoc for this application. The equations employed in the design of the observer are based purely on kinematics, hence they are exact (i.e. they are not affected by uncertainty in the system's parameters) and independent from system's characteristics like the wing area, aerodynamic efficiency, mass, etc. Therefore, the presented experimental results are representative also of systems with larger wing size and different design, including systems with different number of tethers and/or rigid wings. The paper is organized as follows. Section \ref{S:System_description} provides a description of the system, the related model and the considered sensor setup. The sensor fusion algorithms are described in section \ref{S:Approaches} and the experimental results are presented in section \ref{S:results}. A discussion on the results and conclusions are given in section \ref{S:conclusion}.

\section{System description, model equations and sensors setup}\label{S:System_description}
\subsection{System description}\label{SS:yo_yo_descr}
\begin{figure}[hbt]
\centerline{
\includegraphics[bbllx=14mm,bblly=77mm,bburx=175mm,bbury=266mm,width=8cm,clip]{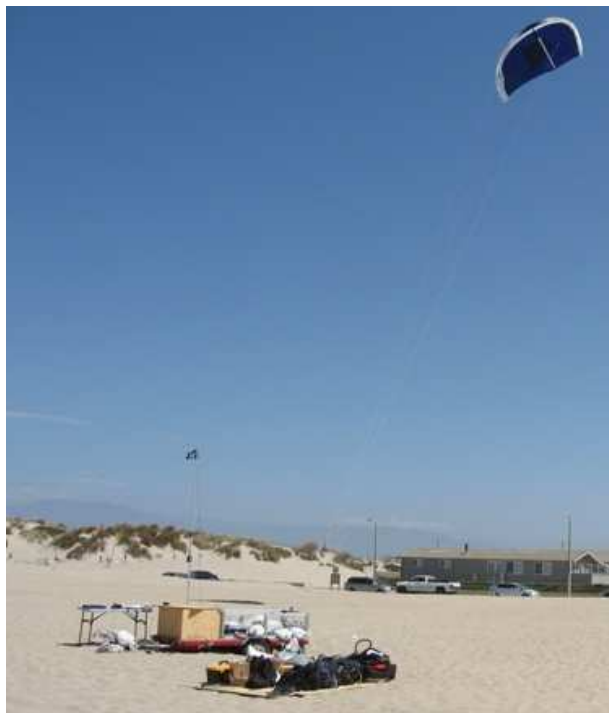}
} \caption{Small-scale prototype for the control of tethered wings built at the University of California, Santa Barbara.}\label{F:proto}
\end{figure}

The considered system is based on a small-scale prototype built at the University of California, Santa Barbara, shown in Fig. \ref{F:proto} (see \cite{Wing_movie} for a short movie clip).
A flexible wing (inflatable power kites are used in the experiments) is linked by three lines to a ground unit (GU). In normal flight conditions, the wing's trajectory evolves downwind with respect to the GU.
The GU is fixed to the ground and it is equipped with actuators, able to influence the wing's path: a human operator or an automatic control algorithm commands the actuators in order to obtain ``figure-eight'' crosswind paths, which maximize the generated forces \cite{FaMi12}. This paper is devoted to the problem of estimating in real-time, in the described flying conditions, the wing's position as well as its ``velocity angle''. In the next section, we provide a formal definition of these variables and we introduce the related notation and model equations.

The prototype operates with a fixed length of the lines of $r=30\,$m, hence it is not able to generate electricity through the synchronized reeling out of the three lines, as it is typical in ground-based airborne wind energy systems (\cite{FaMi12}). Instead, the energy needed for the operation of the system is drawn from batteries installed on the GU. The use of a fixed, short lines' length (as compared with the intended operating conditions of this kind of systems, ranging from 100 to 1000 m of lines' length) does not impair the significance of the presented study, rather it increases it, for ar least four reasons: first, airborne wind energy systems operate with very low line reel-out speed as compared with the other components of the wing's velocity vector, so that the behavior of the wing does not change significantly between fixed line length and variable line length; second, the optimal operation of airborne wind energy generators is obtained with a constant line speed (ideally equal to one third of the wind speed \cite{Loyd80}) and the settings considered here can be regarded to as a particular case of such operating conditions, i.e. with constant line speed equal to zero; third, the use of fixed lines' length yields the largest possible forces for given wind conditions, hence the highest accelerations and angular velocities, making the estimation problem more challenging; finally, the use of short lines implies that the wing's path is contained in a small portion of the aerial space, with consequent much more frequent changes of direction and inversions of motion, again increasing the difficulty of the filtering problem.

\subsection{Model equations}\label{SS:Kine_ME}
\noindent Before stating the model equations used to design our filtering algorithms, we need to introduce a series of right-handed reference systems, as well as the transformations to convert a given vector from one reference to another.
\begin{figure}[hbt]
\centerline{
\includegraphics[bbllx=41mm,bblly=91mm,bburx=237mm,bbury=258mm,width=8cm,clip]{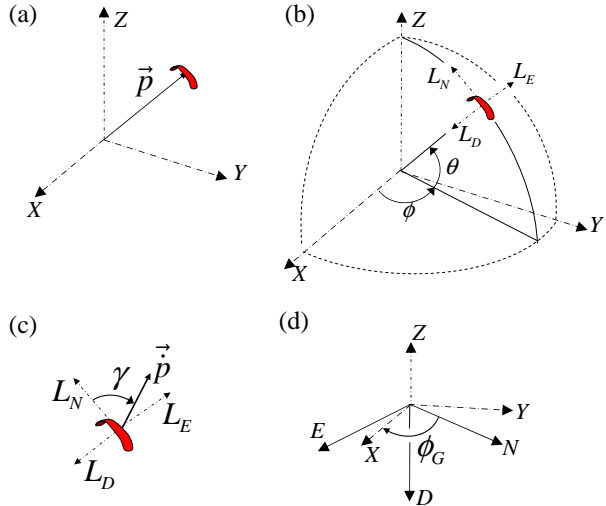}} \caption{(a) Coordinate system $G$ and position of the wing $\vec{p}$; (b) local coordinate system $L$ and spherical coordinates $\theta,\,\phi$;  (c) wing velocity angle $\gamma$; (d) coordinate systems $G$ and $NED$ and angle $\phi_G$.}\label{F:System_frames}
\end{figure}
A first, inertial coordinate system $G\doteq(X,Y,Z)$ (Fig. \ref{F:System_frames}-(a)) is centered at the GU location, with the $X$ axis parallel to the ground, contained in the symmetry plane of the GU and pointing downwind towards the wing. The $Z$ axis is perpendicular to the ground pointing upwards, and the $Y$ axis forms a right-handed system. The wing position vector expressed in the reference system $G$ is denoted by $\vec{p}_G=[p_X,p_Y,p_Z]^T\in\mathbb{R}^3$, where $p_X,\,p_Y$ and $p_Z$ are the scalar components of $\vec{p}_G$ along the axes $X,\,Y$ and $Z$, respectively, and $^T$ stands for the matrix transpose operation. The wing's position can be also expressed in the spherical coordinates $\theta,\,\phi$ (Fig. \ref{F:System_frames}-(b)), where $\theta\in[0,\,\frac{\pi}{2}]$ is the angle between the $(X,\,Y)$ plane and vector $\vec{p}_G$ and $\phi\in[-\pi,\pi]$ is the angle between the $X$ axis and the projection of $\vec{p}_G$ onto the $(X,\,Y)$ plane, taken to be positive for a positive rotation around the $Z$ axis. In particular, we have:
 \begin{equation}\label{E:pos_spherical}
\vec{p}_G=\left[\begin{array}{c}
p_X\\
p_Y\\
p_Z
 \end{array}\right]=r
 \left[\begin{array}{c}
\cos(\theta)\cos(\phi)\\
\cos(\theta)\sin(\phi)\\
\sin(\theta)
 \end{array}\right]
 \end{equation}
where $r$ is the distance between the GU and the wing, equal to the lines' length in our setup. We note that the angle $\theta$ considered here is the complement to $\frac{\pi}{2}$ with respect to the angle employed in previous works (see e.g. \cite{CaFM09c}), and that the $\theta,\,\phi$ angles are different from those used in $\cite{ErSt12}$.

Angles $\theta,\,\phi$ define also  a second, non-inertial Cartesian coordinate system, $L\doteq(L_N,L_E,L_D)$, which we call ``local''.  Vector $\vec{p}_G$ (as well as any other vector) can be expressed in system $L$ by using the rotation matrix $R_{_{G\rightarrow L}}$:
\begin{equation}\label{E:pos_rot}
\vec{p}_L=R_{_{G\rightarrow L}}\vec{p}_G,
\end{equation}\small
\begin{equation}\label{E:R_L_GU}
R_{_{G\rightarrow L}}=	\left[\begin{array}{ccc}
	-\sin{(\theta)} \cos{(\phi)} & -\sin{(\theta)}\sin{(\phi)}&	 \cos{(\theta)}\\
-\sin{(\phi)}&\cos{(\phi)} &0\\ -\cos{(\phi)}\cos{(\theta)}&-\sin{(\phi)}\cos{(\theta)}&-\sin{(\theta)}
\end{array}\right],
\end{equation}\normalsize
and the inverse transformation is given by the rotation matrix $R_{_{L\rightarrow G}}=R_{_{G\rightarrow L}}^{-1}=R_{_{G\rightarrow L}}^T$.
The axes $L_N$ and $L_E$ define the tangent plane at point $\vec{p}_G$ to the sphere of radius $r$, on which the wing's trajectory is confined, and they can be interpreted as local north and east direction relative to such a sphere. Hence, $L_E$ is always parallel to the ground (i.e. to the $(X,Y)$ plane). The axis $L_D$ thus represents the local down, pointing from the wing to the center of the sphere (i.e. to the GU location).

We further consider the wing's coordinate system, $K\doteq(K_x, K_y, K_z)$, and the standard geographical North East Down system, $NED\doteq(N,E,D)$. System $K$ is centered at $\vec{p}_G$, it is  non-inertial and fixed with respect to the wing, i.e. it provides the wing's orientation. In particular, $K_x$ corresponds to the wing's longitudinal symmetry axis, pointing from the trailing to the leading edge, $K_y$ is aligned with the transversal axis of the wing, pointing from the left to the right wing tip, and $K_z$ completes a right handed system. The rotation matrix $R_{_{NED\rightarrow G}}$ is used to express a vector in $NED$-coordinates into $G$-coordinates:
\begin{equation}\label{E:R_NED_{GU}}
	R_{_{NED\rightarrow G}} =
	\begin{pmatrix}
	\cos{(\phi_{G})} & \sin{(\phi_{G})} & 0 \\
	\sin{(\phi_{G})} & -\cos{(\phi_{G})} & 0 \\
	0 & 0 & -1\\
	\end{pmatrix},
\end{equation}
where $\phi_{G}\in[0,2\pi]$ is the angle between the $N$ axis and the $X$ axis, measured by a positive rotation around the $D$ axis (see Fig. \ref{F:System_frames}-(d)). Moreover, a vector expressed in the $K$ system can be converted into the $NED$ system by means of the rotation matrix $R_{_{K\rightarrow NED}}(q)$, where $q=[q_1,q_2,q_3,q_4]^T\in\mathbb{R}^4$ is the quaternion defining the relative orientation between $K$ and $NED$ (see e.g. \cite{La06,CoRo04}):
\begin{equation}\label{E:R_K_NED}
\begin{array}{l}
	R_{_{K\rightarrow NED}}(q) =\\
	\begin{pmatrix}
	2(q_1^2+q_2^2) -1 & 2(q_2q_3-q_1q_4) & 2(q_2q_4+q_1q_3) \\
	2(q_2q_3+q_1q_4) & 2(q_1^2+q_3^2) -1 & 2(q_3q_4-q_1q_2) \\
	2(q_2q_4-q_1q_3) & 2(q_3q_4+q_1q_2) & 2(q_1^2+q_4^2) -1\\
	\end{pmatrix},
\end{array}
\end{equation}
The matrix $R_{_{K\rightarrow G}}\doteq R_{_{NED\rightarrow G}}\,R_{_{K\rightarrow NED}}$ can be used to express a vector in $K$ coordinates into $G$ coordinates.

Finally, we define the wing's velocity angle $\gamma\in[-\pi,\,\pi]$ as follows (see Fig. \ref{F:System_frames}-(c)):
\begin{equation}\label{E:gamma}
	\gamma \doteq \arctan_2{\left(\dot{p}_{L_E},\dot{p}_{L_N}\right)},
\end{equation}
where $\arctan_2{\left(\dot{p}_{L_E},\dot{p}_{L_N}\right)}\in[-\pi,\,\pi]$ is the 4-quadrant arc tangent function and $\dot{p}_{L_E},\,\dot{p}_{L_N}$ are proportional to the sine and cosine of $\gamma$, respectively. The variable $\gamma$ is the angle between the wing's velocity vector and the local north axis, $L_N$, measured positive for a positive rotation around the local down $L_D$. We note that, since the considered system has fixed line length, vector $\vec{\dot{p}}$ is always contained in the $(L_N,\,L_E)$ plane, however the definition \eqref{E:gamma} is more general and  holds also in case of variable line length. Angle $\gamma$ is particularly important as feedback variable for automatic control algorithms (see e.g. \cite{FZMK13_arxiv}), since it can be easily linked to the wing's path: as an example, $\gamma=0$, $\gamma=\frac{\pi}{2}$ and $\gamma=\pi$ indicate that the wing is moving, respectively, towards the local north, parallel to the ground towards the local east, or towards the ground.

\begin{figure}[!hbt]
\centering
\includegraphics[bbllx=19mm,bblly=200mm,bburx=267mm,bbury=267mm,width=8cm,clip]{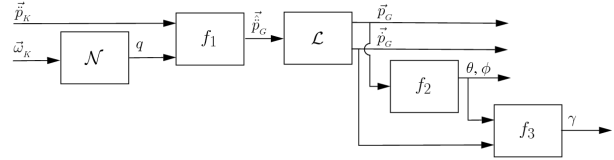}
\caption{Scheme of the considered kinematic model.}
\label{F:System_kine}
\end{figure}

We are now in position to introduce the model equations that we consider in this work. These equations are based entirely on the system's kinematics, i.e. they are basically obtained by differentiating twice the vector $\vec{p}_G(t)$ with respect to the continuous time variable $t$. Kinematic equations bring two important advantages: 1) they provide an exact model, i.e. there are no neglected dynamics even for flexible wings, and 2) they do not depend on any of the system's characteristics, like mass, shape, moments of inertia and aerodynamics of the wing. Indeed these features influence the motion of the wing  through complex, infinite dimensional nonlinear dynamics, whose inputs are the steering command given by the control system and the (unmeasured) wind, yet such dynamics are totally irrelevant for our scope if a measure or estimate of the acceleration $\vec{\ddot{p}}_G(t)$, which is the input to our model, is available. However, while the kinematics in the inertial frame $G$ are given by linear operators (i.e. derivatives), we still end up with nonlinear model equations, due to the fact that we measure the involved variables in different reference frames. In order to facilitate the observer design, we split the model into five interconnected subsystems, as shown in Fig. \ref{F:System_kine}. The involved variables are the wing position, velocity and acceleration in the inertial reference $G$, $\vec{p}_G(t),\,\vec{\dot{p}}_G(t),\,\vec{\ddot{p}}_G(t)$, the wing acceleration in the non-inertial reference $K,\,\vec{\ddot{p}}_K(t)$, the quaternion $q(t)$, finally the wing angular velocity  in the reference $K$, $\vec{\omega}_K(t)$.  Referring to Fig. \ref{F:System_kine}, the model equations are:
\begin{subequations}\label{E:kine_model}
\begin{align}
	 \mathcal{L} :&\; \begin{aligned}
      	\begin{bmatrix} \dfrac{d\vec{p}_{G}(t)}{dt} \\ \dfrac{d\vec{\dot{p}}_{G}(t)}{dt}\end{bmatrix} & =
	\begin{bmatrix} 0_3 & I_3 \\ 0_3 & 0_3 \end{bmatrix} \begin{bmatrix} \vec{p}_{G}(t) \\ \vec{\dot{p}}_{G}(t)\end{bmatrix}
	+ \begin{bmatrix} 0_3 \\ I_3 \end{bmatrix} \vec{\ddot{p}}_G(t)
       \end{aligned}\label{E:kine_linear}\\\,&\nonumber\\
	 \mathcal{N}  :&\;\begin{aligned}
	\dot{q}(t) = \frac{1}{2}
		\begin{bmatrix}
		0 & -\omega_{K_x} & -\omega_{K_y} & -\omega_{K_z}\\
		\omega_{K_x} & 0 & -\omega_{K_z} & \omega_{K_y}\\
		\omega_{K_y} & \omega_{K_z} & 0 & -\omega_{Kx}\\
		\omega_{K_z} & -\omega_{K_y} & \omega_{K_x} & 0\\
		\end{bmatrix}
		\begin{bmatrix} q_1 \\ q_2 \\ q_3 \\ q_4 \end{bmatrix}
     	\end{aligned}\label{E:kine_nonlinear2}\\\,&\nonumber\\
     	 f_1 :&\; \begin{aligned}
       \vec{\ddot{p}}_{G}(t)=R_{_{NED\rightarrow G}}(\phi_G) R_{_{K\rightarrow NED}}(q(t)) \;  \vec{\ddot{p}}_{K}(t)
       \end{aligned}\label{E:kine_nonlinearf1}\\\,&\nonumber\\
	 f_2:&\;\left\{
	 \begin{array}{l}\theta(t)=\arcsin{\left(\frac{p_Z(t)}{r}\right)}\\
\phi(t)=\arctan_2{\left(p_Y(t),p_X(t)\right)}
\end{array}\right.
     	\label{E:kine_nonlinearf2}\\\,&\nonumber\\
     	 f_3:&\; \begin{aligned}
	\left\{\begin{array}{l}\vec{\dot{p}}_L=R_{_{G\rightarrow L}}\vec{\dot{p}}_G\\
\gamma(t)=\arctan_2{\left(\dot{p}_{L_E}(t),\dot{p}_{L_N}(t)\right)}\end{array}\right.\end{aligned}\label{E:kine_nonlinearf3}
\end{align}
\end{subequations}
where $0_3$ is a 3$\times$3 matrix of zeros, and $I_3$ is the 3$\times$3 identity matrix.
Thus, the considered model has $\vec{\ddot{p}}_{K},\,\vec{\omega}_K(t)$ as inputs and  $\vec{p}_G(t),\,\vec{\dot{p}}_G(t),\,q(t)$ as states, and it is composed by a linear dynamical system, $\mathcal{L}$, a nonlinear one, $\mathcal{N}$, and three static nonlinear functions,  $f_1,\,f_2,\,f_3$. The only involved parameters are the constant angle $\phi_G$, giving the orientation of the GU with respect to the geographical North, and the length $r$ of the lines. Both these parameters can be measured accurately and are assumed to be known exactly here. In the next section, we will describe the different considered sensor setups and indicate the corresponding measured variables.

\subsection{Sensors setup}\label{SS:sensors}
\begin{figure}[hbt]
\centerline{
\includegraphics[bbllx=13mm,bblly=77mm,bburx=266mm,bbury=267mm,width=8cm,clip]{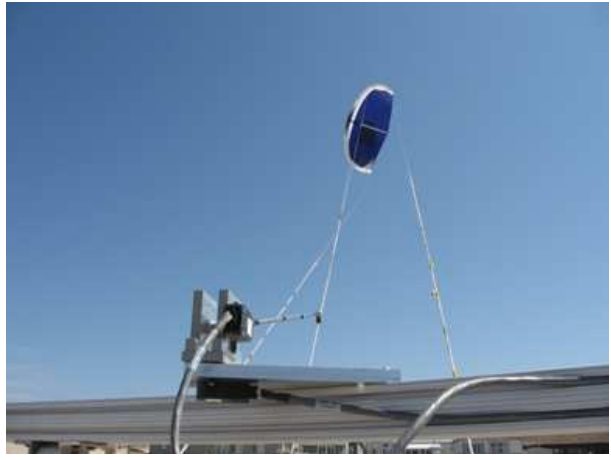}
} \caption{Line angle sensor}\label{F:power_line_encoder}
\end{figure}

\begin{figure}[htb]
\centerline{
\begin{tabular}{c}
(a) \\
\includegraphics[bbllx=32mm,bblly=103mm,bburx=152mm,bbury=284mm,width=5cm,clip]{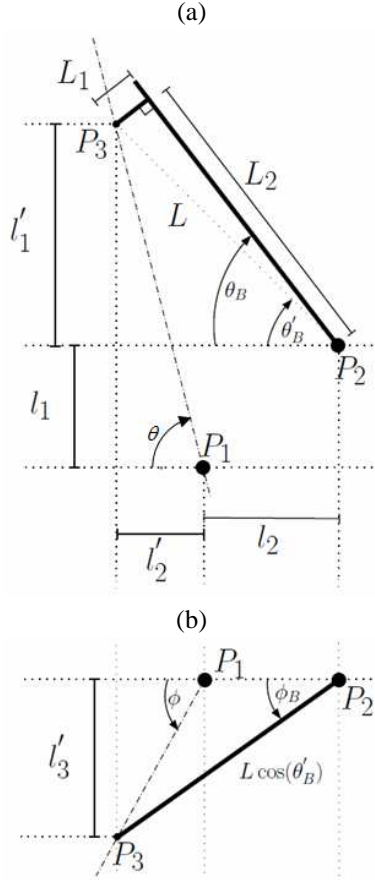}\\
(b) \\
\includegraphics[bbllx=32mm,bblly=15mm,bburx=152mm,bbury=94mm,width=5cm,clip]{encoder_draw_new2.eps}
\end{tabular}}
\caption{Diagram of the line angle sensor: (a) side view with $\phi=0$ and (b) top view. Dash-dotted line: center line of the wing; thick solid lines: metal rod and pulley linking the encoders to the center line.}
\label{F:line_sensor}
\end{figure}

Several sensors are installed on the GU and onboard the wing, giving a range of different possibilities for the design of observers for $\vec{p}_G(t)$ and $\gamma(t)$. A GPS and three magnetometers are installed on the GU, providing its geographical location in $NED$ as well as the angle $\phi_G$. Moreover, the GU is equipped with a line angle measurement system (shown in Fig. \ref{F:power_line_encoder}),
which provides a direct measurement of the angle between the main line connecting the wing to the GU and the axes of the inertial reference system $G$. This sensor has been developed and built ad-hoc for this application and it employs two incremental encoders. The encoders are connected to the center line via a metal rod and a small pulley which allows the center line to slide with low friction with respect to the rod while the wing is flying. The first encoder measures the angle between the metal rod and the $(X,Y)$ plane, while the second one measures the angle between the $X$ axis and the projection of the metal rod onto $(X,Y)$. The incremental encoders provide accurate absolute angle measurements through a quite standard integration algorithm that sums the encoder's counts, after computing their sign (i.e. the direction of motion) based on two quadrature signals shifted by 90$^\circ$ in phase. The obtained angle measurements are very robust against noise and drift, since they are based on discrete impulses.  Moreover we used an additional index impulse to re-set a standard position each time the encoders went through it. Driven by the movement of the center line, the metal rod rotates following the wing's path, and the encoders measure such rotations in terms of the above-mentioned angles. More specifically, the  readings from the two incremental encoders, denoted as $\theta_B$ and $\phi_B$, can be converted into measurements of $\theta$ and $\phi$ via the following equations (see Fig. \ref{F:line_sensor}):
\begin{subequations} \label{E:enc_side}
\begin{align}
     	L		&	=\sqrt{L^2_1 + L^2_2} \label{E:enc_side_a} \\
   \theta^{'}_B 	&	 =\theta_B-\arctan\left(\frac{L_1}{L_2}\right) \label{E:enc_side_b} \\
     	l^{'}_1	&	=L\sin{(\theta^{'}_B)} \label{E:enc_side_c} \\
        l^{'}_2	&	 =L\cos{(\theta^{'}_B)}\cos{(\phi_B)}-l_2\label{E:enc_side_d}\\
        l^{'}_3	&	=L\cos{(\theta^{'}_B)}\sin{(\phi_B)} \label{E:enc_top_a}\\
	\theta	&	 =\arctan{\left(\frac{l^{'}_1+l_1}{\sqrt{l^{'2}_2+l^{'2}_3}}\right)} \label{E:enc_side_e}\\
	\phi	&	=\arctan_2{\left(l^{'}_3,\,l^{'}_2\right)}, \label{E:enc_top_c}
\end{align}
\end{subequations}
\noindent where $L_1$ and $L_2$ are, respectively, the lengths of the metal rod and the pulley on top of it, and $l_1,\,l_2$ are the distances between the encoders' position $P_2$ and the attachment point of the center line on the GU, $P_1$. The quantities $L_1,\,L_2,\,l_1,\,l_2$ are fixed and known. Each encoder has a resolution of $\frac{2\pi}{400}\,$rad, i.e. 400 counts for each full revolution. With the considered line length of $30\,$m, this translates in about 0.2$\,$m of resolution in each direction.

As to the onboard sensors, the wing is equipped with an Inertial Measurement Unit (IMU) manufactured by SBG Systems$^\circledR$, comprising  three accelerometers, three gyroscopes, three magnetometers, a barometer, and a GPS. A 900 MHz radio is also installed on the wing and transmits the measurements collected by the IMU to a receiver on the GU, at a rate of 115200 Baud. The inertial sensors of the IMU are calibrated after production. The accelerometers have a bandwidth of 50$\,$Hz, a measurement range of $\pm5\,$g (where g is the gravity constant), a bias of $\pm4\,10^{-3}\,$g, nonlinearity $<\,10^{-2}\,$g, and noise density of $2.5\times10^{-4}$g$/\sqrt{\text{Hz}}$. The gyroscopes have a bandwidth of 40$\,$Hz, a measurement range of $\pm300\,^{\circ}/$s, a bias of $\pm\,10^{-1\,\circ}/$s, nonlinearity  $<3\,10^{-1\,\circ}/$s, and noise density of $5\,10^{-2\,\circ}/($s$\sqrt{\text{Hz}})$. The magnetometers have a bandwidth of 500$\,$Hz, a measurement range of $\pm1.2\,\text{Gauss}$, a bias of $\pm5\,10^{-4}\,$Gauss, nonlinearity $<2.4\,10^{-3}\,$Gauss, and noise density of $1\,10^{-5}\times\text{Gauss}/\sqrt{\text{Hz}}$. The barometer provides a measurement of barometric altitude at 9$\,$Hz with a resolution of $0.2\,$m. The mentioned sensors are suitable for the considered application, where the main motion components lie in the range 0.15-1$\,$Hz, angular velocities are contained in the range $\pm200\,^{\circ}/$s, and accelerations lie in the range $\pm4.5\,$g.\\
The signals of all sensors mentioned so far are acquired with a sampling time of $T_s=0.02\,$s to be used in the estimation algorithms.

The onboard GPS has a nominal horizontal accuracy (i.e. on the $(N,E)$ plane) of $\pm2.5\,$m and a sampling time of $0.25\,$s. The delay between the GPS readings and the measurements given by the other sensors on the IMU is compensated by synchronizing the time stamps of each signal, which are also available.

Finally, an anemometer is also installed on the GU and provides wind speed and direction measurements at 4$\,$m above the ground. Although the wind measurements are not directly used in the filtering algorithms, they are useful to analyze the obtained results, as we show in section \ref{S:results}.

The complete measurement, estimation and control system was implemented on a real-time machine by SpeedGoat$^\circledR$, programmed using MatLab$^\circledR$ and xPC Target$^\circledR$.

In the following, we implicitly assume that:
\begin{itemize}
  \item the line angle sensor provides a measure of the $\theta,\,\phi$ angles;
  \item the  GPS provides a measure of the position $\vec{p}_G$;
  \item the barometer provides a measure of $p_Z$;
  \item the IMU is fixed with respect to the kite, hence the accelerometers, gyroscopes and magnetometers provide measurement of, respectively, $\vec{\ddot{p}}_K$, $\vec{\omega}_K$, and $q$.
\end{itemize}
These assumptions imply that the wing's lines are parallel to its position vector $\vec{p}_G$, that the position of the IMU coincides with $\vec{p}_G$ and that the IMU itself is fixed to the wing, so that the measured accelerations and angular speeds corresponds to those of the wing. Given the short length of the wing's lines and high forces developed in crosswind motion (typically ranging from 500$\,$N to 3000$\,$N in our experiments), the hypotheses above are reasonable in the considered application. We provide some considerations on the validity of such assumptions in the presence of longer lines in the discussion in section \ref{S:conclusion}.

\section{Sensor fusion algorithms}\label{S:Approaches}

In this section we present different algorithms to estimate the wing's position and its velocity angle, $\vec{p}_G(t)$ and $\gamma(t)$. In the following, we use the notation  `$\,\tilde{\,}\,$' to indicate  noise-corrupted measurements and unfiltered estimates of a given variable, e.g. $\vec{\tilde{p}}_G$ is the measurement of the position vector seen from system $G$, $\vec{p}_G$. Moreover, we indicate with `$\,\hat{\,}\,$' the estimated variables, i.e. the outputs of the observer.

As a first step, we exploit the structure of the considered kinematic model to separate the problem of estimating the wing orientation (i.e. the quaternion $q(t)$) from the one of estimating its position and velocity. In fact, since the nonlinear system $\mathcal{N}$ is known and the IMU provides a measurement of both its input and its state, i.e. $\vec{\tilde{\omega}}_K(t)$ and $\tilde{q}(t)$, an extended Kalman filter (EKF) approach can be employed to obtain the value of $\hat{q}(t)$. The use of EKFs for the absolute orientation of rigid bodies using gyroscopes and magnetometers is well-studied in the literature (see e.g. \cite{Saba11}), and most commercial IMUs have already an EKF implemented. In particular, the IMU employed in this study already includes its own EKF for this purpose, and the filter has been calibrated on a vibrational and rotational testbed after production. The obtained attitude estimate have a range of $360^\circ$ around all three axes $(K_x, K_y, K_z)$ with static accuracies of $\pm0.5^\circ$ around $K_x$ and $K_y$ and  $\pm1^\circ$ around $K_z$, a dynamic accuracy of $\pm1^\circ\,$RMS, resolution $<5\,10^{-2\,\circ}$ and repeatability error $<2\,10^{-1\,\circ}$. Such performance are sufficient to capture the rotational motion of the wing during crosswind flight with high accuracy. We indicate the EKF filter as $\mathcal{N}_{EKF}$.

With the estimate $\hat{q}(t)$ provided by $\mathcal{N}_{EKF}$, we can then compute the rotation matrix $R_{K\rightarrow NED}(\hat{q}(t))$ by using \eqref{E:R_K_NED}. Then, we compute an estimate $\vec{\hat{\ddot{p}}}_{G}(t)$ of $\vec{\ddot{p}}_{G}(t)$ through the static nonlinear function $\hat{f}_1$ given by:
\begin{equation}\label{E:acc_estim}
\hat{f}_1:\vec{\hat{\ddot{p}}}_{G}(t)=R_{_{NED\rightarrow G}}(\phi_G) R_{_{K\rightarrow NED}}(\hat{q})\vec{\tilde{\ddot{p}}}_{K}(t)+\left[\begin{array}{c}0\\0\\\text{g}\end{array}\right].
\end{equation}
$\hat{f}_1$ is obtained from $f_1$ \eqref{E:kine_nonlinearf1} by using the estimated rotation matrix $R_{K\rightarrow NED}(\hat{q}(t))$ instead of the ``true'' one, and then by removing the bias given by the gravity acceleration g. The latter is present in our setup due to the characteristics of the employed MEMS accelerometers. The approach described so far is used as a preliminary step for any of the observers described in the following sections, whose goal is to estimate the wing's position and velocity. In particular, consider that the estimate $\vec{\hat{\ddot{p}}}_{G}(t)$ can be re-written as:
\begin{equation}\label{E:acc_noise}
\vec{\hat{\ddot{p}}}_{G}(t)=\vec{\ddot{p}}_{G}(t) + \vec{\eta}_p(t),
\end{equation}
where $\vec{\eta}_p(t)\in\mathbb{R}^3$ is some estimation error that we consider as process noise. Then, the estimate $\vec{\hat{\ddot{p}}}_{G}(t)$ can be regarded to as a noise-corrupted input for the linear system $\mathcal{L}$ (see Fig. \ref{F:System_kine}). Following this idea, we will use a steady-state Kalman filter, based on $\mathcal{L}$, to compute $\vec{\hat{p}}_G(k)$ and $\vec{\hat{\dot{p}}}_G(k)$. In order to do so, we discretize $\mathcal{L}$ by forward difference with the employed sampling time $T_s$:
\begin{equation}\label{E:sys_equ1}
\begin{array}{l}
	\begin{bmatrix} \vec{p}_G(k+1) \\ \vec{\dot{p}}_G(k+1) \end{bmatrix} =\\\\
	\underbrace{	\begin{bmatrix} I_3 & T_s\,I_3 \\ 0_3 & I_3\end{bmatrix} }_A 		 \begin{bmatrix} \vec{p}_G(k) \\ \vec{\dot{p}}_G(k) \end{bmatrix} +	 \underbrace{ 	 \begin{bmatrix} 0_3 \\ T_sI_3  \end{bmatrix} }_B  			 \left(\vec{\ddot{p}}_{G}(k) + \vec{\eta}_p(k)\right)\\
	\vec{\tilde{p}}_G(k) = \underbrace{	\begin{bmatrix} I_3 & 0_3\end{bmatrix}}_C 		 \begin{bmatrix} \vec{p}_G(k) \\ \vec{\dot{p}}_G(k) \end{bmatrix} + \vec{\eta}_m(k),
\end{array}
\end{equation}
where $k\in\mathbb{Z}$ is the discrete time variable and $\vec{\eta}_m\in\mathbb{R}^3$ represents the output measurement noise. In classical Kalman filtering theory, $\vec{\eta}_p \text{ and } \vec{\eta}_m$  are assumed to be independent white Gaussian processes  with covariance matrices $Q$ and $R$, respectively \cite{WaBi01}. Here, the measurement and process noises can be reasonably assumed to be independent, since they pertain to completely different sensors (accelerometers, gyroscopes and magnetometers for the input $\vec{\hat{\ddot{p}}}_{G}(k)$ and either GPS and barometer or line angle sensor for the output $\vec{\tilde{p}}_G(k)$), yet they are not white Gaussian processes, since the measurements are the result of algorithms like $\mathcal{N}_{EKF}$ or the GPS. However, the matrices $Q$ and $R$ can still be tuned on the basis of the characteristics of the employed sensors by following rather simple guidelines, described in section \ref{SS:tuning}.

\begin{figure}[htb]
\centerline{
\begin{tabular}{c}
(a) \\
\includegraphics[bbllx=15mm,bblly=197mm,bburx=264mm,bbury=266mm,width=8.7cm,clip]{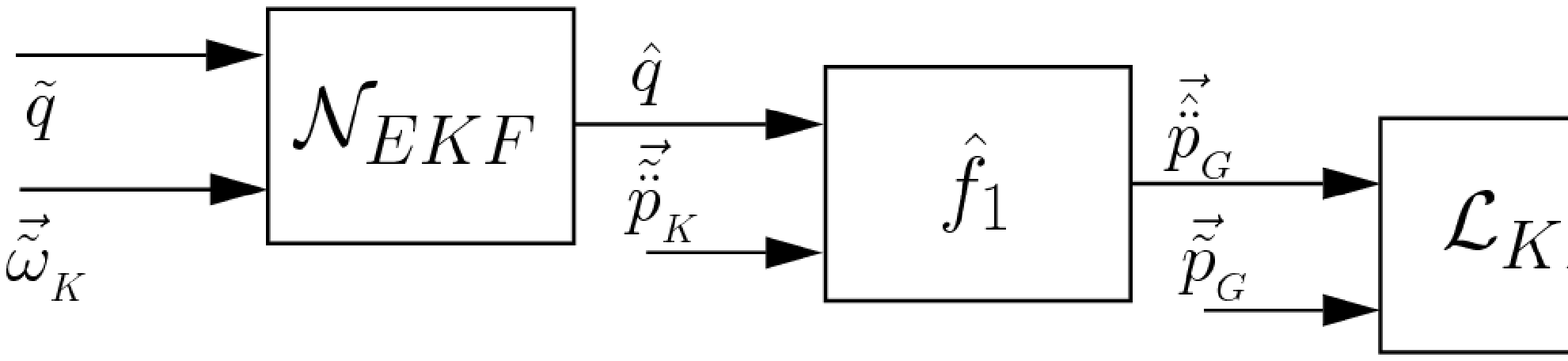} \\
(b) \\
\includegraphics[bbllx=15mm,bblly=179mm,bburx=264mm,bbury=266mm,width=8.7cm,clip]{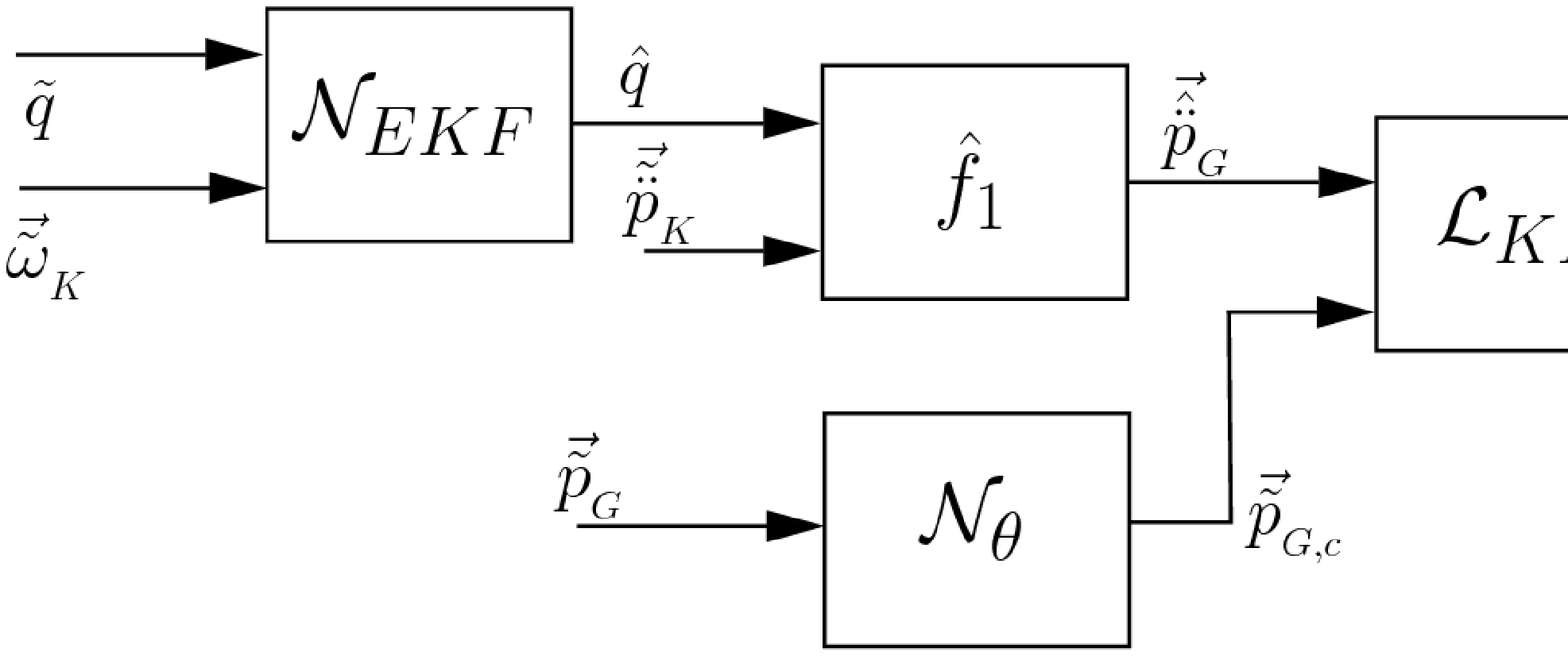} \\
(c) \\
\includegraphics[bbllx=15mm,bblly=179mm,bburx=264mm,bbury=266mm,width=8.7cm,clip]{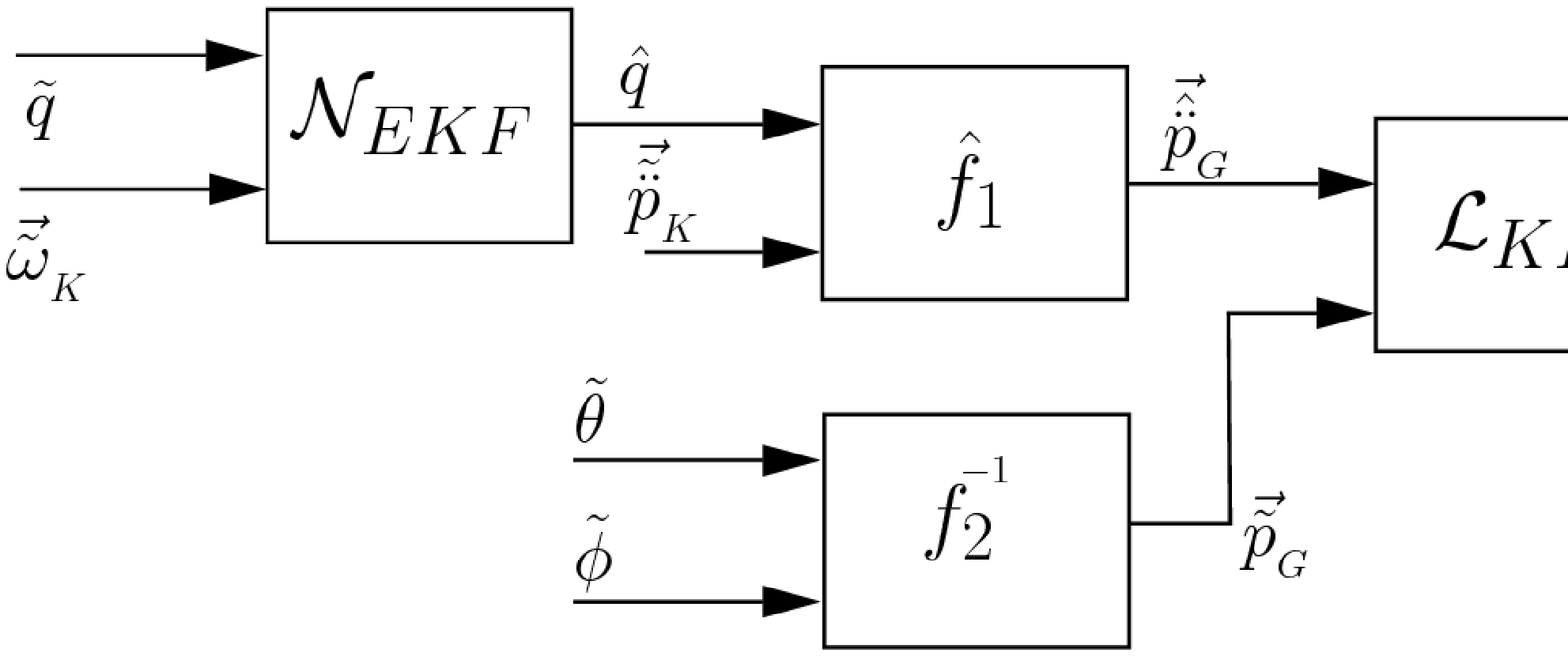}
\end{tabular}}
\caption{Schemes of the designed observers for the first, second and third approach ((a), (b) and (c), respectively).}
\label{F:filters}
\end{figure}

\noindent The Kalman Filter is then obtained by the following equations \cite{WaBi01,BaLiKi93}:\\
Time update:\\
\begin{equation}\label{E:time_update}
	\begin{array}{rcl}
	\begin{bmatrix} \vec{\hat{p}}^-_G(k) \\ \vec{\hat{\dot{p}}}^-_G(k) \end{bmatrix} & = & A \begin{bmatrix} \vec{\hat{p}}_G(k-1) \\ \vec{\hat{\dot{p}}}_G(k-1) \end{bmatrix} + B\vec{\hat{\ddot{p}}}_G(k) \\
	\end{array}
\end{equation}
Measurement update:\\
\begin{equation}\label{E:meas_update}
	\begin{array}{ccc}
	\begin{bmatrix} \vec{\hat{p}}_G(k) \\ \vec{\hat{\dot{p}}}_G(k) \end{bmatrix} & = & \begin{bmatrix} \vec{\hat{p}}^-_G(k) \\ \vec{\hat{\dot{p}}}^-_G(k) \end{bmatrix} + K\left(\vec{\tilde{p}}_G(k) - C \begin{bmatrix} \vec{\hat{p}}^-_G(k) \\ \vec{\hat{\dot{p}}}^-_G(k) \end{bmatrix}\right) \\
	\end{array}
\end{equation}
where $\vec{\hat{p}}^-_G(k),\,\vec{\hat{\dot{p}}}^-_G(k)$ are the a-priori state estimates at step $k$ given the knowledge of the process prior to step $k$, while $\vec{\hat{p}}_G(k),\,\vec{\hat{\dot{p}}}_G(k)$ are the a-posteriori state estimates at step $k$ given the knowledge of the measurement $\vec{\tilde{p}}_G(k)$. The steady-state Kalman gain $K$ is computed off-line as:
\begin{equation}\label{K_gain}
	K = AP_{\infty}C^T (CP_{\infty}C^T + R)^{-1},
\end{equation}
where $P_{\infty}$ satisfies the following Algebraic Riccati Equation:
\begin{equation}\label{P}
\begin{array}{l}
	P_{\infty} =\\ AP_{\infty}A^T -AP_{\infty}C^T(CP_{\infty}C^T+R)^{-1}CP_{\infty}A^T + BQB^T.
\end{array}
\end{equation}
We will describe next the three different approaches considered in this study. All three of them employ the described procedure to obtain $\vec{\hat{\ddot{p}}}_G(k)$ and then the Kalman filter \eqref{E:time_update}-\eqref{E:meas_update} to estimate position and velocity, but they use different strategies and sensors to obtain the position measurement $\vec{\tilde{p}}_G(k)$. A scheme of each of the three approaches is shown in Fig. \ref{F:filters}, where we indicate the Kalman filter as $\mathcal{L}_{KF}$.

\subsection{First approach: GPS and barometer}\label{SS:1st_appr}

We design a first observer (Fig. \ref{F:filters}-(a)) by using the GPS to obtain the measurements $\tilde{p}_X,\,\tilde{p}_Y$, and the barometer for $\tilde{p}_Z$. Hence, the two sensors together provide the position measurement $\vec{\tilde{p}}_G(k)$ to be used with the Kalman filter \eqref{E:time_update}-\eqref{E:meas_update}. While the GPS can also provide a measurement of the altitude, the related error (about $\pm$50 m in our setup) is too large to be used in the considered application. The barometer, on the other hand, is quite accurate provided that an initial tuning procedure is carried out for each test to remove the bias induced by the weather conditions. The GPS measurements are obtained by computing the difference of the readings between the onboard GPS and the one placed on the GU. Since the accelerometers have a sampling frequency of $50\,$Hz, while the GPS provides measurements at a $4\,$Hz rate, we use a multi-rate Kalman filter in this first approach, which consists in performing the time update (\ref{E:time_update}) at each sampling time of $T_s=0.02\,$s, and the measurement update (\ref{E:meas_update}) every $0.25\,$s, i.e. when the new GPS measurement is available.

\subsection{Second approach: GPS and barometer with geometric correction}\label{SS:2nd_appr}
The filter designed in the second approach employs the same sensors as the previous one, but it also carries out a correction of the variables $\tilde{p}_X,\,\tilde{p}_Y$ given by the GPS, in order to project the measured position on the sphere of radius $r$. This correction is based on two observations: 1) the wing motion is constrained on such a sphere, 2) the measurements $\tilde{p}_X,\,\tilde{p}_Y$, given by the GPS, are less reliable than the measurement $\tilde{p}_Z$ provided by the barometer. Given $\tilde{Z}$, by using \eqref{E:kine_nonlinearf2} we compute a measure $\tilde{\theta}$ of angle $\theta$, then, we correct $\tilde{p}_X,\,\tilde{p}_Y$ by re-scaling them in order to match with the line length $r$ projected onto $(X,\,Y)$, i.e. $r\cos(\tilde{\theta})$ (compare \eqref{E:pos_spherical} and Fig. \ref{F:System_frames}-(b)):
\begin{equation}\label{E:GPS_corr}
\begin{array}{ccc}
	\tilde{p}_{X,\text{c}}&=& \tilde{p}_X\frac{r\cos(\tilde{\theta})}{\sqrt{\tilde{p}_X^2+\tilde{p}_Y^2}}\\
\tilde{p}_{Y,\text{c}}&=& \tilde{p}_Y\frac{r\cos(\tilde{\theta})}{\sqrt{\tilde{p}_X^2+\tilde{p}_Y^2}}
\end{array}
\end{equation}
Then, we use the vector $\vec{\tilde{p}}_{G,c}=[\tilde{p}_{X,\text{c}},\,\tilde{p}_{Y,\text{c}},\,\tilde{p}_Z]^T$ as position measurement for the multi-rate Kalman filter. This second approach is shown in Fig. \ref{F:filters}-(b), where the correction \eqref{E:GPS_corr} is indicated as $\mathcal{N}_\theta$.

\subsection{Third approach: line angle sensor}\label{SS:3rd_appr}

The third observer we consider (Fig. \ref{F:filters}-(c)) employs the line angle sensor to obtain the position measurement. Hence, in this approach we move away from using the GPS and barometer measurements. By using \eqref{E:kine_nonlinearf2} we have:
\begin{equation}\label{E:line_enc_conv}
\begin{array}{ccc}
	\tilde{p}_X(k) &=& r \cos(\tilde{\theta}(k)) \cos(\tilde{\phi}(k))\\
	\tilde{p}_Y(k) &=& r \cos(\tilde{\theta}(k)) \sin(\tilde{\phi}(k))\\
	\tilde{p}_Z(k) &=& r \sin(\tilde{\theta}(k)),
\end{array}
\end{equation}
where $\tilde{\theta}(k),\,\tilde{\phi}(k)$ are the measurements provided at $50\,$Hz by the line angle sensor. Due to the use of incremental encoders, the measurement noise on $\tilde{\theta}(k),\,\tilde{\phi}(k)$ is basically given just by the quantization error resulting from the encoders' resolution (see section \ref{SS:sensors}).\\
$\,$\\

The three approaches presented above differ by two main aspects: the use of different sensors for the position measurements, which differentiate the first and second approaches (GPS and barometer) from the third one (line angle), and the inclusion of the known kinematic constraint given by the tether, which forces the wing to move on a sphere of radius $r$. The latter aspect is not considered in the first approach, while it is accounted for in the second one (where we project the measured position on the sphere of radius $r$ before doing the measurement update in $\mathcal{L}_{KF}$) and in the third one (where we directly compute the position from the spherical coordinates, hence automatically obtaining a measured position that lies on the sphere). It can be noted that we do not include the tether constraint directly in the model \eqref{E:kine_model}, rather we enforce it on the measured output $\vec{\tilde{p}}_{G}$. Adding the constraint to \eqref{E:kine_model} would still yield an exact model (i.e. without model uncertainty), at least as long as the straight line assumption holds (see section \ref{S:conclusion} for some discussion in this regard). However, the filtering algorithm would then have to take such constraints into account, hence standard linear techniques like the Kalman filter could not be used anymore, and ad-hoc modifications would have to be implemented. We tried one such approach by projecting the state on the constraint for the time update \eqref{E:time_update}, before implementing the measurement update \eqref{E:meas_update}. This strategy did not yield any advantage with respect to the algorithms presented above for approaches 2 and 3, nor did  a combination of the two techniques, i.e. projecting the state on the constraint in \eqref{E:time_update} and then use projected measurements in \eqref{E:meas_update}. In summary, the inclusion of the tether constraint in the model used in $\mathcal{L}_{KF}$ does not yield any advantage w.r.t. enforcing the constraint on the position measurements and using a ``free particle'' linear model in $\mathcal{L}_{KF}$. The latter approach, which we presented here, has the advantage of simplicity and well-understood theory for the stability of the estimation error dynamics.

\subsection{\(\gamma\) filter}\label{SS:gamma_filter}
In order to obtain an estimate of the velocity angle $\gamma$ \eqref{E:gamma}, we first use \eqref{E:kine_nonlinearf2} to obtain the angles $\hat{\theta},\,\hat{\phi}$ from the filtered position $\vec{\hat{p}}_G$. The latter is obtained, together with the velocity estimate $\vec{\hat{\dot{p}}}_G$, using one of the observers presented in the previous sections. Then, we estimate the rotation matrix $R_{_{G\rightarrow L}}(\hat{\theta},\,\hat{\phi})$ by using \eqref{E:R_L_GU}, and we compute an estimate of the velocity vector in the $L$ frame, $\vec{\hat{\dot{p}}}_L$, as:
\begin{equation}\label{pdot_local}
\vec{\hat{\dot{p}}}_L=R_{_{G\rightarrow L}}(\hat{\theta},\,\hat{\phi})\vec{\hat{\dot{p}}}_G.
\end{equation}
Using $\vec{\hat{\dot{p}}}_L$, we compute an unfiltered variable $\tilde{\gamma}\simeq\gamma$ as (using the definition \eqref{E:gamma}):
\begin{equation}\label{E:gamma_tilde}
\tilde{\gamma}=\arctan_2{\left(\hat{\dot{p}}_{L_E},\hat{\dot{p}}_{L_N}\right)},
\end{equation}
Then, we employ a standard Luenberger observer \cite{Ellis02} to obtain a filtered version of $\tilde{\gamma}$, based on the following state-space model:
\[
\begin{array}{rcl}
	\begin{bmatrix} \gamma(k+1) \\ \dot{\gamma}(k+1) \end{bmatrix} &=&
	\begin{bmatrix} 1 & T_s \\ 0 & 1\end{bmatrix} 		 \begin{bmatrix} \gamma(k) \\ \dot{\gamma}(k) \end{bmatrix}
\end{array}
\]
In particular, the state-space equations of the Luenberger observer are:
\begin{equation}\label{E:gamma_filter}
\begin{array}{rcl}
	\begin{bmatrix} \hat{\gamma}(k+1) \\ \hat{\dot{\gamma}}(k+1) \end{bmatrix} &=&
	\begin{bmatrix} 1 & T_s \\ 0 & 1\end{bmatrix} 		 \begin{bmatrix} \hat{\gamma}(k) \\ \hat{\dot{\gamma}}(k) \end{bmatrix}+K_\gamma\left(\tilde{\gamma}(k)-\hat{\gamma}(k)\right).
\end{array}
\end{equation}
Hence, the observer yields a filtered estimate of the velocity angle $\hat{\gamma}$, as well as of its rate $\hat{\dot{\gamma}}$. The former is required in order to have a smooth signal, suitable to be used for feedback control, the latter can be  used as a feedback variable, too, and it is useful to study the turning dynamics of the wing in response to a steering deviation (see e.g. \cite{FZMK13_arxiv}). The only tuning parameter for the Luenberger observer is its static gain $K_\gamma\in\mathbb{R}^2$: in section \ref{SS:tuning}, we provide guidelines on how to choose this gain in light of the considered application.\\
$\,$

Clearly, the three different approaches presented above to estimate the wing's position and velocity yield estimates of $\gamma$ with different accuracy. In sections \ref{SS:results_1_2} and \ref{SS:results_3}, we discuss the performance obtained with the considered filters in experimental tests.

\section{Tuning guidelines and experimental results}\label{S:results}
We designed three different observers for $\vec{\hat{p}}_G$, according to the approaches described in section \ref{S:Approaches}, and consequently three observers for $\hat{\gamma}$. Throughout this section, we use the superscripts $^1,\,^2,\,^3$ to indicate the filtered position  obtained with the approach of section \ref{SS:1st_appr}, \ref{SS:2nd_appr} and \ref{SS:3rd_appr}, respectively. We employed different wings in the experiments, in order to demonstrate that the proposed approaches are independent from system's features like wing area, mass, efficiency, etc. In particular,  we used three different Airush One$^\circledR$ inflatable power kites, with 12$\,$m$^2$, 9$\,$m$^2$ and 6$\,$m$^2$ area. The length of the lines is $r=30\,$m.

Before presenting the results of the experimental tests, we provide some considerations on the tuning of the algorithms described in section \ref{S:Approaches}.

\subsection{Tuning guidelines}\label{SS:tuning}
For a chosen sampling time (i.e. $T_s=0.02\,$s in our case), the parameters to be tuned in the presented approaches are three: the matrices $Q$ and $R$ that define the gain $K$ of $\mathcal{L}_{KF}$ \eqref{E:time_update}-\eqref{E:meas_update}, and the gain $K_\gamma$ of the Luenberger observer \eqref{E:gamma_filter}.

\begin{figure}[hbt]
\centerline{
\includegraphics[bbllx=27mm,bblly=72mm,bburx=178mm,bbury=198mm,width=7.5cm,clip]{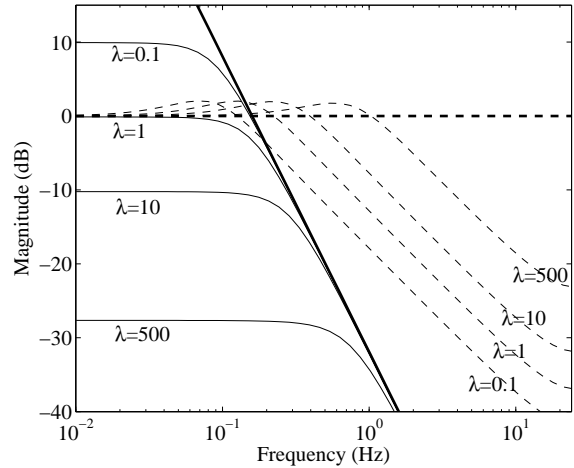}} \caption{Bode plot of the magnitude of the discrete-time transfer functions $F^{\text{KF}}_u(z)\doteq\hat{p}_X(z)/\hat{\ddot{p}}_X(z)$ (solid lines) and $F^{\text{KF}}_y\doteq\hat{p}_X(z)/\tilde{p}_X(z)$ (dashed lines) for different values of weighting ratio $\lambda\doteq Q/R$, together with their respective asymptotic behaviors (thick lines).}\label{F:KF_TF}
\end{figure}

\emph{Kalman filter}. For the choice of $Q$ and $R$, a  rather simple analysis can be made in this specific case. In fact, in the inertial frame the position components $p_X,\,p_Y$ and $p_Z$ are the outputs of three double integrators that are decoupled (compare \eqref{E:kine_linear}). Hence, if $Q$ and $R$ are chosen to be diagonal, i.e.:
\[
	R =
	\begin{bmatrix}
	R_{11} & 0 & 0\\
	0 & R_{22} & 0\\
	0 & 0 & R_{33}\\
	\end{bmatrix} \quad
	Q =
	\begin{bmatrix}
	Q_{11} & 0 & 0\\
	0 & Q_{22} & 0\\
	0 & 0 & Q_{33}\\
	\end{bmatrix},
\]
then the filter $\mathcal{L}_{KF}$ enjoys a similar structure, so that one can carry out the analysis by considering just one of the three directions, and the considerations hold also for the other two components. Let us consider for example the $X$ direction. The related system has the acceleration $\ddot{p}_X$ as input, the position $p_X$ as output, and the gain of the observer is determined by the ratio $\lambda=Q_{11}/R_{11}$, which is the only tuning parameter. We can then analyze the transfer functions of $\mathcal{L}_{KF}$ from each of its two inputs, i.e. the acceleration $\hat{\ddot{p}}_X$ and the position measure $\tilde{p}_X$, to its output, for example the filtered position $\hat{p}_X$:
\[
\hat{p}_X(z)=F^{\text{KF}}_u(z)\hat{\ddot{p}}_X(z)+F^{\text{KF}}_y(z)\tilde{p}_X(z),
\]
where $x(z)$ is the $z$-transform of the discrete-time signal $x(k)$. Fig. \ref{F:KF_TF} shows the Bode plot of the magnitude of the discrete-time transfer functions $F^{\text{KF}}_u$ and $F^{\text{KF}}_y$ as a function of $\lambda$. Both functions are low-pass filters, with similar bandwidth. Let us first consider function $F^{\text{KF}}_y$. For ``low'' frequencies,
this function is close to 1 (showed as thick dashed line in Fig. \ref{F:KF_TF}) irrespective of $\lambda$, as one would expect since the filter output $\hat{p}_X$ and the measurement $\tilde{p}_X$ correspond to the same physical quantity. As regards function  $F^{\text{KF}}_u(z)$, it can be noted that for ``high'' frequencies its magnitude is the same as that of function $\frac{T_s^2}{(z-1)^2}$ (showed as thick solid line in Fig. \ref{F:KF_TF}), irrespective of $\lambda$. This is also expected, since the latter function corresponds to a double integrator and indeed the position $p_X$ is obtained (less the initial conditions) by integrating twice the acceleration (i.e. the input of $F^{\text{KF}}_u$). The frequency ranges where these equivalences hold true depend on $\lambda$: the smaller this value, the larger the frequency range where $F^{\text{KF}}_u$ corresponds to a double integrator and the smaller the range where $F^{\text{KF}}_y\approx1$, and vice-versa. A similar analysis can be done for the transfer functions from the two inputs to the estimated velocity $\hat{\dot{p}}_X$.
The considerations above, combined with the characteristics of the employed sensors, provide quite intuitive guidelines on how to tune $\lambda$ (hence $Q$ and $R$). If the sensors used to measure the position have good
bandwidth and low high-frequency noise, a higher value of $\lambda$ should be chosen, hence relying on the position sensors for a larger range of low frequencies, and on the integration of the acceleration for higher frequencies. This is the case of the line angle sensor in approach 3, for which we set $\lambda=500$ (see Fig. \ref{F:KF_TF}). If, on the contrary, the employed position sensor has poor dynamic performance, a lower value of $\lambda$ can be used, in order to try to rely on the integration of the acceleration also for low and mid-range frequencies. This is the case of the GPS of approaches 1 and 2, for which we
chose $\lambda=10$. Clearly, these guidelines do not have an absolute validity, since they have to be applied by considering the relative quality of the employed sensors.

\emph{Luenberger observer}. The observer \eqref{E:gamma_filter} is a dynamical system with one input, i.e. the unfiltered velocity angle $\tilde{\gamma}$ \eqref{E:gamma_tilde}, and two outputs, i.e. the filtered velocity angle $\hat{\gamma}$ and its derivative $\hat{\dot{\gamma}}$. By changing the gain $K_\gamma$, different bandwidths and zero-pole locations of the transfer functions between the input and each of the outputs can be obtained.
\begin{figure}[hbt]
\centerline{
\includegraphics[bbllx=27mm,bblly=78mm,bburx=179mm,bbury=199mm,width=8cm,clip]{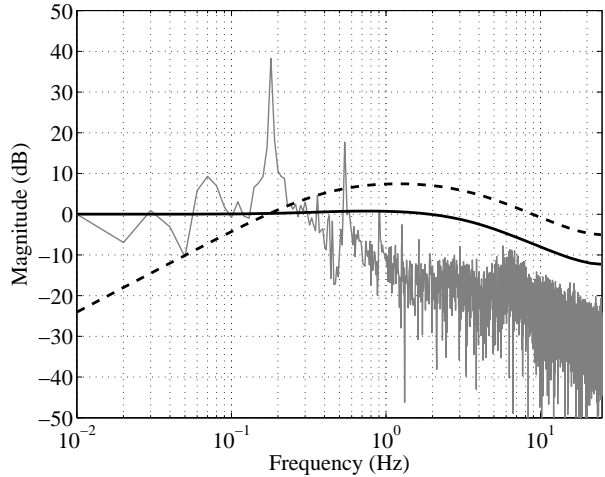}} \caption{Bode plot of the magnitude of the discrete-time transfer functions $F^{\text{LO}}_{y_1}(z)\doteq \hat{\gamma}(z)/\tilde{\gamma}(z)$ (solid lines) and $F^{\text{LO}}_{y_2}(z)\doteq \hat{\dot{\gamma}}(z)/\tilde{\gamma}(z)$ (dashed lines) obtained with the observer's gain $K_\gamma=[0.4\;0.9]^T$, and magnitude of the fast Fourier transform of the input signal $\tilde{\gamma}(z)$ (gray line).}\label{F:LO_TF}
\end{figure}
Fig. \ref{F:LO_TF} shows the Bode plot of the magnitude of  transfer functions $F^{\text{LO}}_{y_1}(z)\doteq \hat{\gamma}(z)/\tilde{\gamma}(z)$ and $F^{\text{LO}}_{y_2}(z)\doteq \hat{\dot{\gamma}}(z)/\tilde{\gamma}(z)$, obtained with the gain $K_\gamma=[0.4\;0.9]^T$ that we chose for our tests. The same figure also shows the magnitude of the fast Fourier transform of the signal $\tilde{\gamma}(z)$ (i.e. the input to the observer), computed from the experimental data collected during more than four hours of autonomous flight in crosswind conditions. From the latter, the main frequency component can be easily identified at around 0.16$\,$Hz, corresponding to the frequency of a full figure-eight path (whose period is typically around 5-6$\,$s in our experimental setup, see e.g Fig. \ref{F:gamma_w_wo_IMU} in section \ref{SS:results_3}). Another peak at around 0.54$\,$Hz can be also noted, as well as a third one at 0.9$\,$Hz.  As a general guideline, the gain $K_\gamma$ shall be chosen such that the important components of the input signal are not distorted in magnitude nor in phase. With the chosen value of $K_\gamma=[0.4\;0.9]^T$, the transfer function $F^{\text{LO}}_{y_1}(z)$ is such that the frequency components up to around 1.5$\,$Hz pass the filter with little magnitude distortion and phase lag (see Figure \ref{F:LO_TF}, solid line). As to function $F^{\text{LO}}_{y_2}(z)$, it can be noted that in a similar range of frequencies this filter behaves like the discrete-time derivative $\frac{z-1}{T_s}$, hence providing an estimate of the turning rate $\dot{\gamma}$ (Fig. \ref{F:LO_TF}, dashed line). Both filters then attenuate signal components with frequencies >2-3 Hz.

\subsection{Experimental results: first and second approaches}\label{SS:results_1_2}
The third approach resulted to be the most accurate one, thanks to the use of the line angle sensor, which practically yields exact position readings. Hence, in this section we take the estimates $\vec{\hat{p}}_G^3$ and $\hat{\gamma}^3$, obtained with the third approach, as ``true values'', and we evaluate the performance of the first two approaches with respect to the third one. Referring to section \ref{SS:tuning}, we employed $\lambda=10$ for all three motion directions, and $K_\gamma=[0.4,\,0.9]^T$.
\begin{figure}[hbt]
\centerline{
\includegraphics[bbllx=8mm,bblly=93mm,bburx=132mm,bbury=199mm,width=7.5cm,clip]{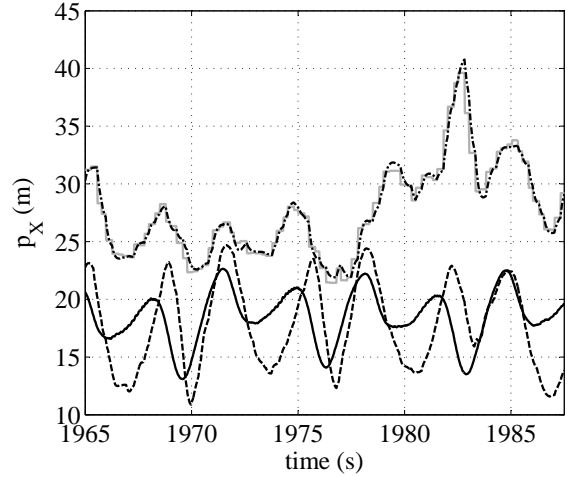}}
\caption{Experimental results. Estimates of the position along the $X$ axis, $p_X(t)$, obtained with the first (dash-dotted line), second (dashed) and third (solid) approaches, and raw data provided by the GPS (gray line).} \label{F:est_pos_x_direc}
\end{figure}
\begin{figure}[hbt]
\centerline{
\includegraphics[bbllx=28mm,bblly=76mm,bburx=180mm,bbury=197mm,width=7.5cm,clip]{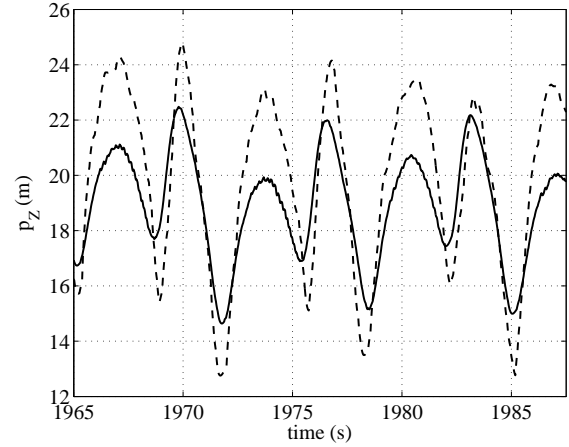}}
\caption{Experimental results. Estimates of the position along the $Z$ axis, $p_Z(t)$, obtained with either the first and second (dashed line) or the third (solid) approach.} \label{F:est_pos_z_direc}
\end{figure}
An example of the courses of the position components along the $X$ 
and $Z$ axes are shown in Figs. \ref{F:est_pos_x_direc}-\ref{F:est_pos_z_direc}.
It can be noted that the error between $\hat{p}_X^1,\,\hat{p}_X^2$ and $\hat{p}_X^3$ can be quite high (of the order of 10 m), with $\hat{p}_X^2$ being generally better than $\hat{p}_X^1$.  This shows that the correction \eqref{E:GPS_corr} that we introduced in the second approach indeed improves the estimate with respect to using the plain GPS reading (compare the dash-dotted and dashed lines in Fig. \ref{F:est_pos_x_direc}). Moreover, both $\hat{p}_X^1$ and $\hat{p}_X^2$ are affected by some delay with respect to $\hat{p}_X^3$.
Such a delay, together with the quite poor accuracy on the $(X,Y)$ plane, give rise to significant errors in the $\gamma$ estimates. Similar results are obtained for the $Y$ axis, while the estimates $\hat{p}_Z^1,\,\hat{p}_Z^2$ result to be more accurate and they are not affected by delays, thanks to the use of the barometric altitude instead of the GPS, see Fig. \ref{F:est_pos_z_direc}. We note that for the component $p_Z$, the first two approaches are equivalent, hence resulting in the same filtered values. The  considerations above are confirmed by the analysis of extensive experimental data. Table \ref{T:RMSE} shows the root mean square errors (RMSE) between the first two approaches and the third one, for all of the estimated variables, grouped in different ranges of wind speeds measured at 4$\,$m from the ground with the anemometer. The results of Table \ref{T:RMSE} account for about 5 hours of operation at different wind speeds. Considering that a single figure-eight loop has a period of 5-6 seconds, the data in the table correspond to about 3300 complete figure-eight paths.
\begin{table}
\centering
\caption{Experimental results. Root mean square errors between the estimates obtained with either the first or second approach and the third one, for different wind speeds measured at 4$\,$m above the ground.}
\resizebox{7.5cm}{!} {\begin{tabular}[!hbt]{|l|l|l|l|l|l|}
\hline \highertop
Wind speed (m/s)&<2 & 2-3 & 3-4 & >4 \\ \hline\highertop
$p_X^1$ (m)& 7.73& 14.11& 16.36& 46.23\\\highertop
$p_X^2$ (m)& 5.05& 7.03& 9.12& 13.63\\\hline\highertop
$p_Y^1$ (m)& 10.32& 12.81& 15.40& 24.53\\\highertop
$p_Y^2$ (m)& 10.10& 10.15& 10.33& 10.65\\\hline\highertop
$p_Z^1$ (m)& 1.68& 2.98& 4.27& 7.70\\\highertop
$p_Z^2$ (m)& 1.68& 2.98& 4.27& 7.70\\\hline\highertop
$\gamma^1$ (rad)& 1.35& 1.44& 1.57& 2.24\\\highertop
$\gamma^2$ (rad)& 1.44& 1.47& 1.79& 2.34\\\hline
\end{tabular}\label{T:RMSE} }
\end{table}
The results in the Table confirm that the second approach is generally better than the first one in estimating the wing's position, especially at larger wind speeds (see Table \ref{T:RMSE} for 3 m/s and above), while the two approaches give similar errors on the $\gamma$ estimates, which deviate significantly from the third approach. Moreover, it can be noted that the accuracy of the first two approaches gets worse as the wind speed increases. Since the wing's speed depends linearly on the wind speed (see e.g. \cite{FaMP11}), this results indicate that the GPS performance get worse as the wing movements get faster and changes of direction are more frequent. Indeed the provided GPS accuracy ($\pm$2.5$\,$m in the $(N,E)$ plane) is valid for steady state conditions only. Finally, it can be noted that also the estimates of $p_Z$ get generally worse with larger wind speed. The reason for this result is that the barometric altitude reading is influenced by the change of air pressure due to the increased wing speed and hence increased airflow on the barometer, which is tuned in static conditions. This phenomenon can be compensated by implementing a correction e.g. based on the wing's speed or line force, which are related to the airflow.

\subsection{Experimental results: third approach}\label{SS:results_3}

\begin{figure}[hbt]
\centerline{
\includegraphics[bbllx=26mm,bblly=75mm,bburx=180mm,bbury=200mm,width=7.5cm,clip]{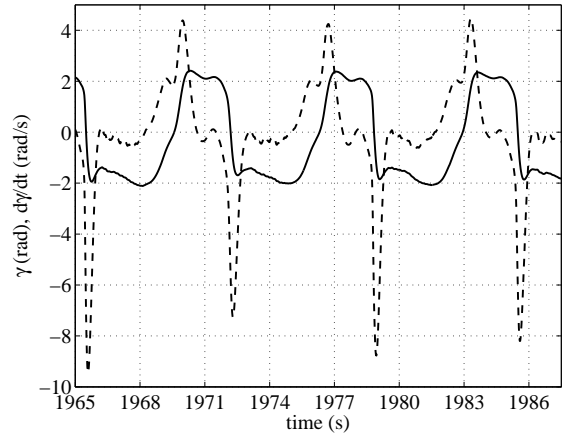}} \caption{Time courses of the filtered velocity angle $\hat{\gamma}$ (solid line), and its derivative $\hat{\dot{\gamma}}$ (dashed line). Design parameters: $\lambda=500$, $K_\gamma=[0.4,\,0.9]^T$. Wing size: 12 m$^2$, wind speed: 2.3 m/s at 4$\,$m from the ground.}\label{F:Filter_gamma}
\end{figure}

\begin{figure*}[!htb]
\centerline{
\begin{tabular}{cc}
\multicolumn{2}{c}{(a)}\\
\includegraphics[bbllx=25mm,bblly=76mm,bburx=180mm,bbury=199mm,width=7cm,clip]{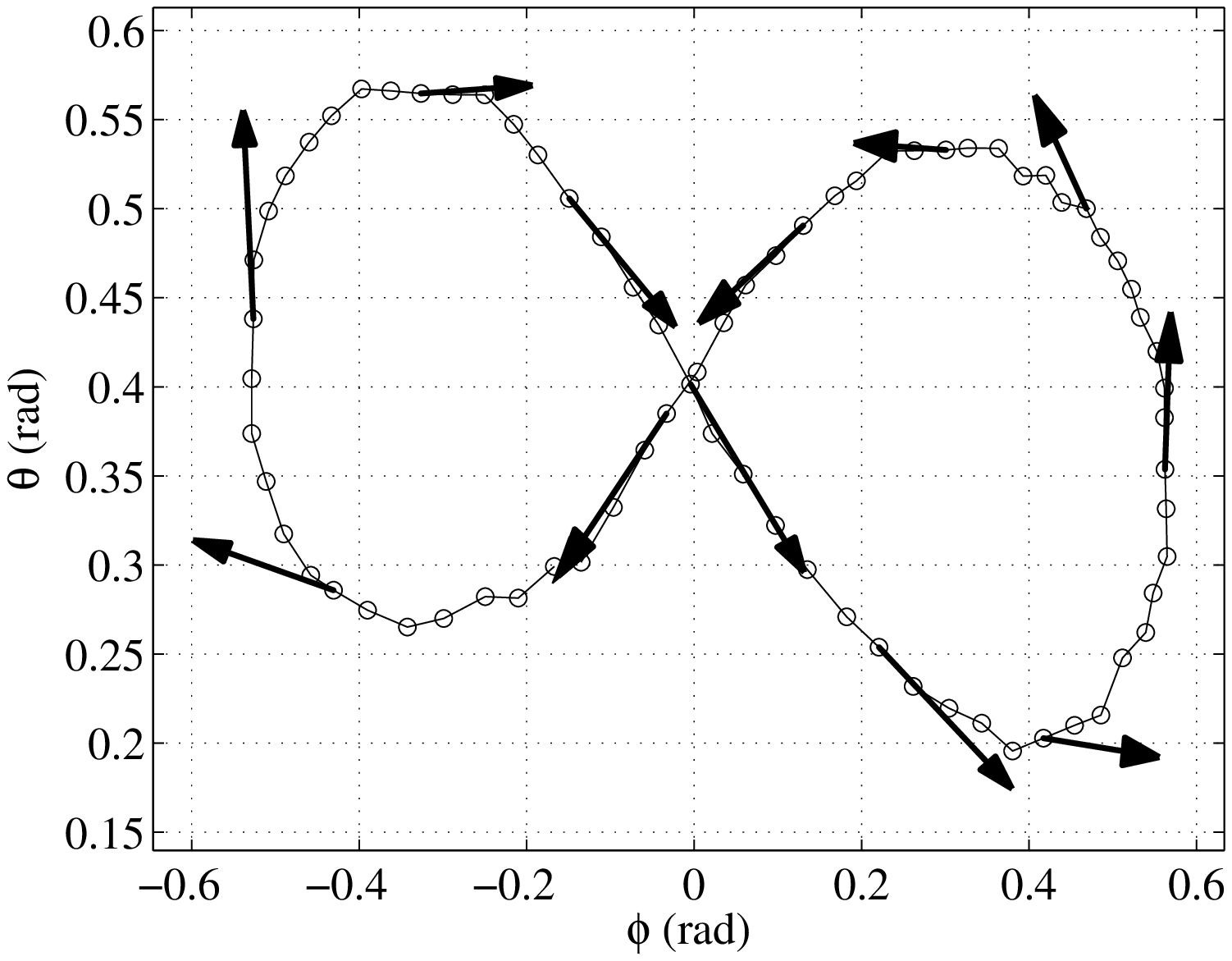} &
\includegraphics[bbllx=25mm,bblly=76mm,bburx=180mm,bbury=199mm,width=7cm,clip]{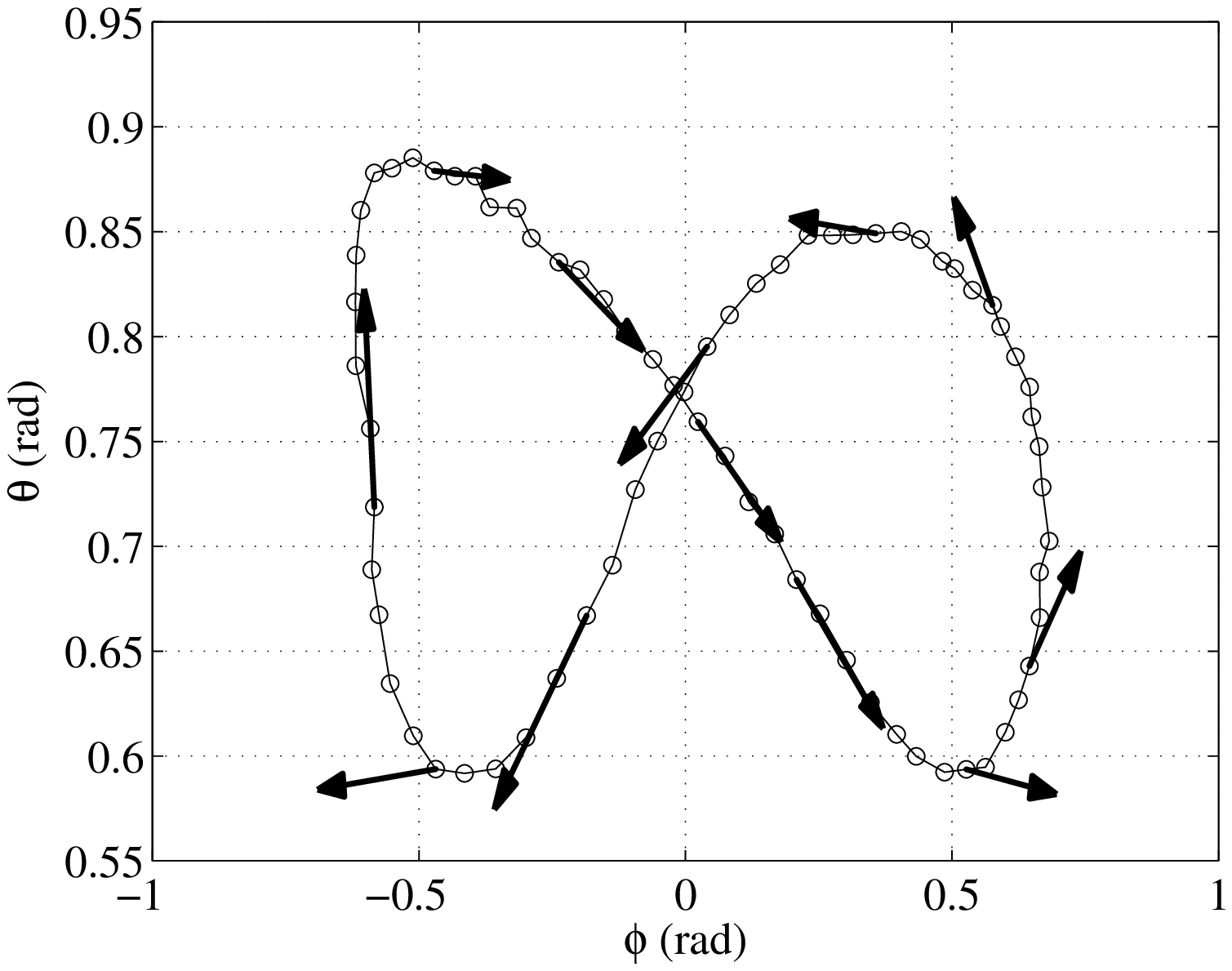} \\
\multicolumn{2}{c}{(b)}\\
\includegraphics[bbllx=25mm,bblly=76mm,bburx=180mm,bbury=199mm,width=7cm,clip]{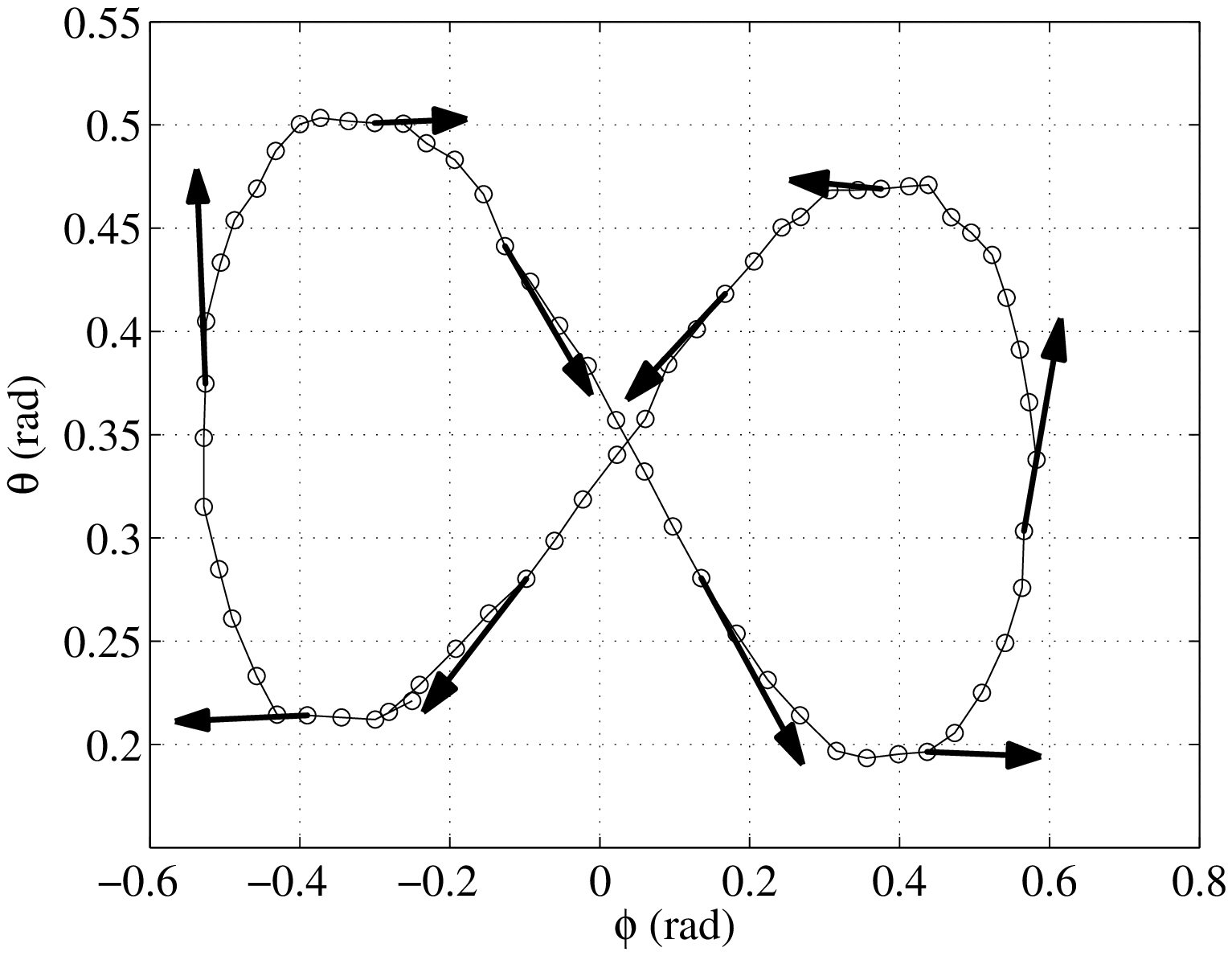} &
\includegraphics[bbllx=25mm,bblly=76mm,bburx=180mm,bbury=199mm,width=7cm,clip]{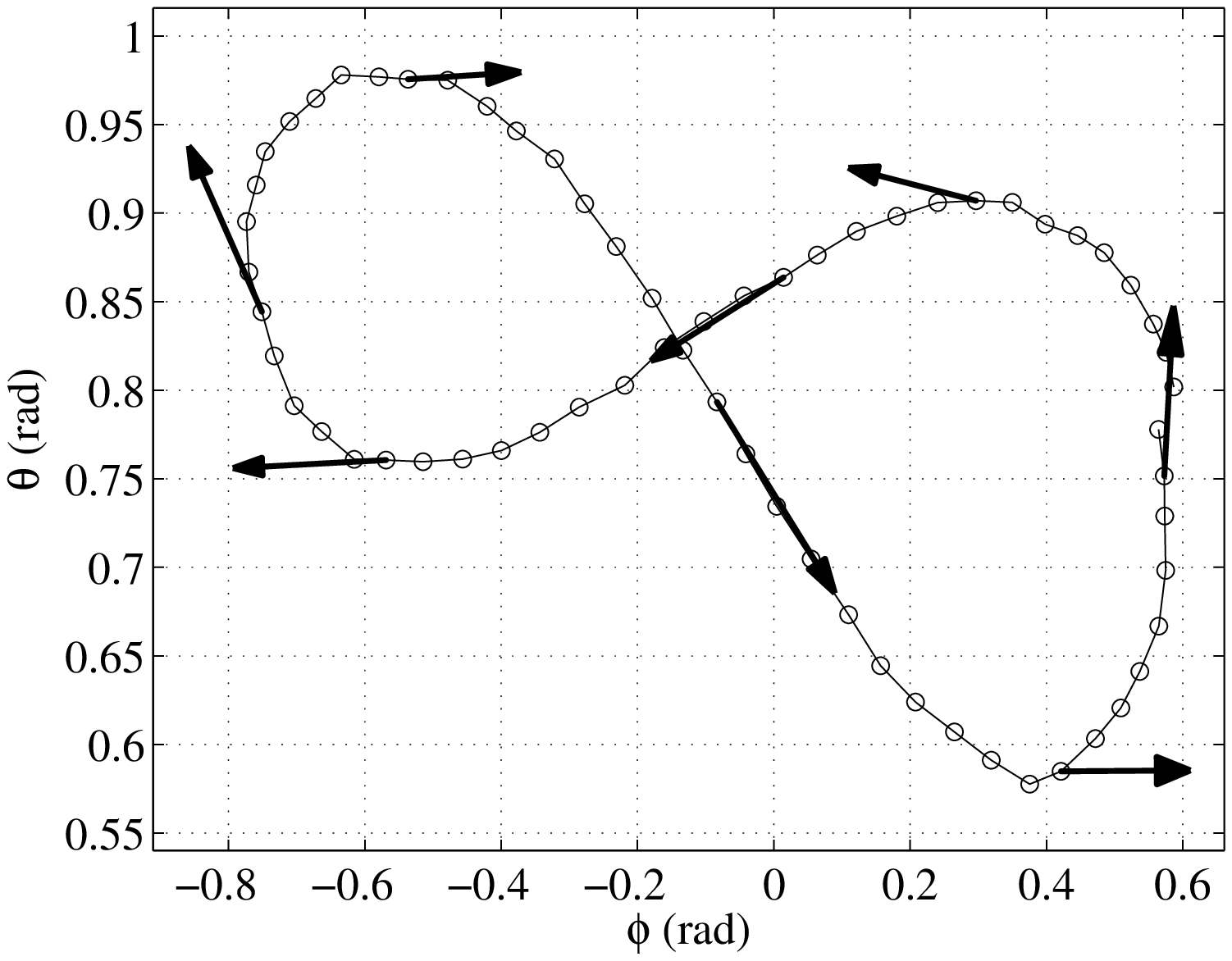}\\
\multicolumn{2}{c}{(c)}\\
\includegraphics[bbllx=25mm,bblly=76mm,bburx=180mm,bbury=199mm,width=7cm,clip]{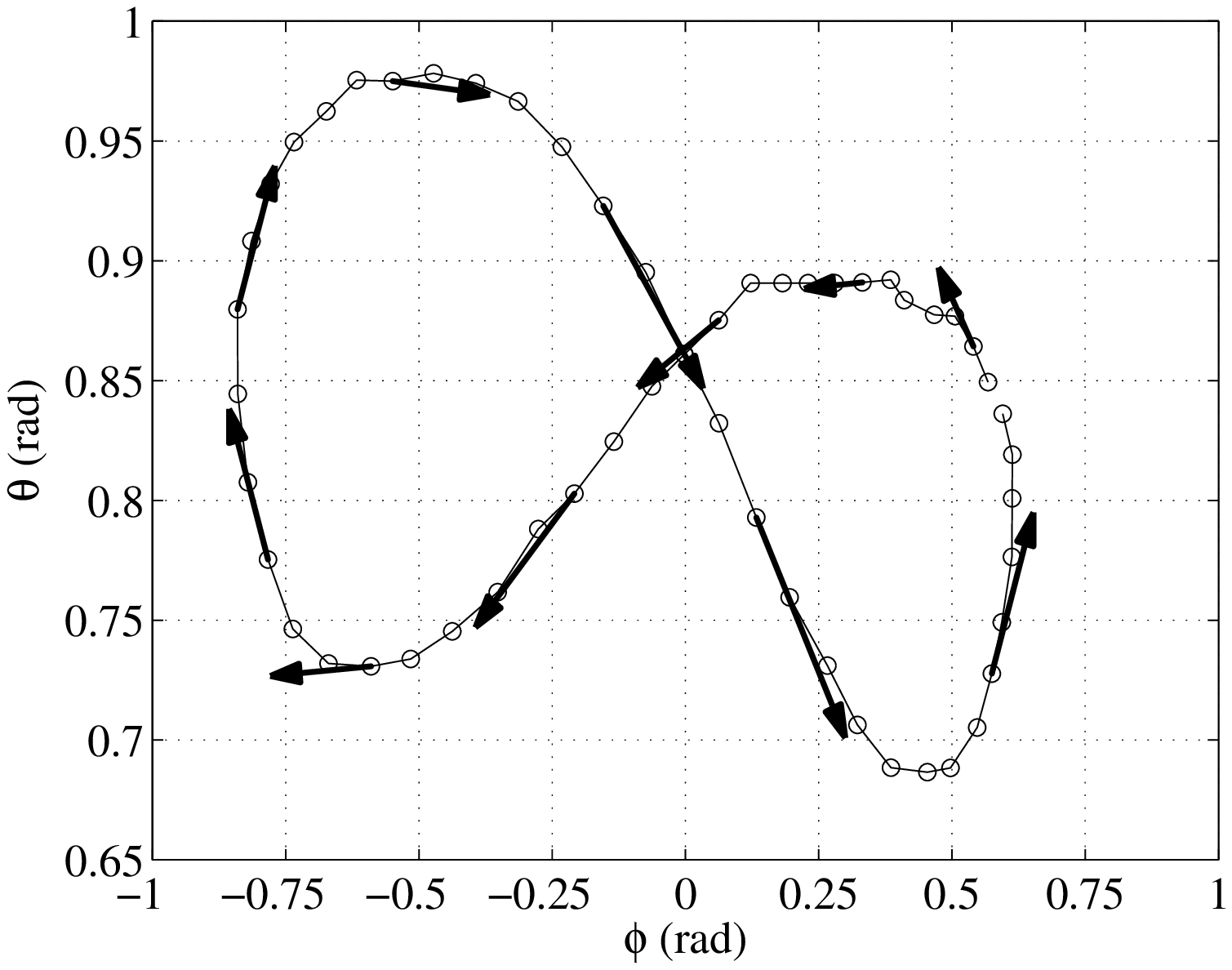}
&
\includegraphics[bbllx=25mm,bblly=76mm,bburx=180mm,bbury=199mm,width=7cm,clip]{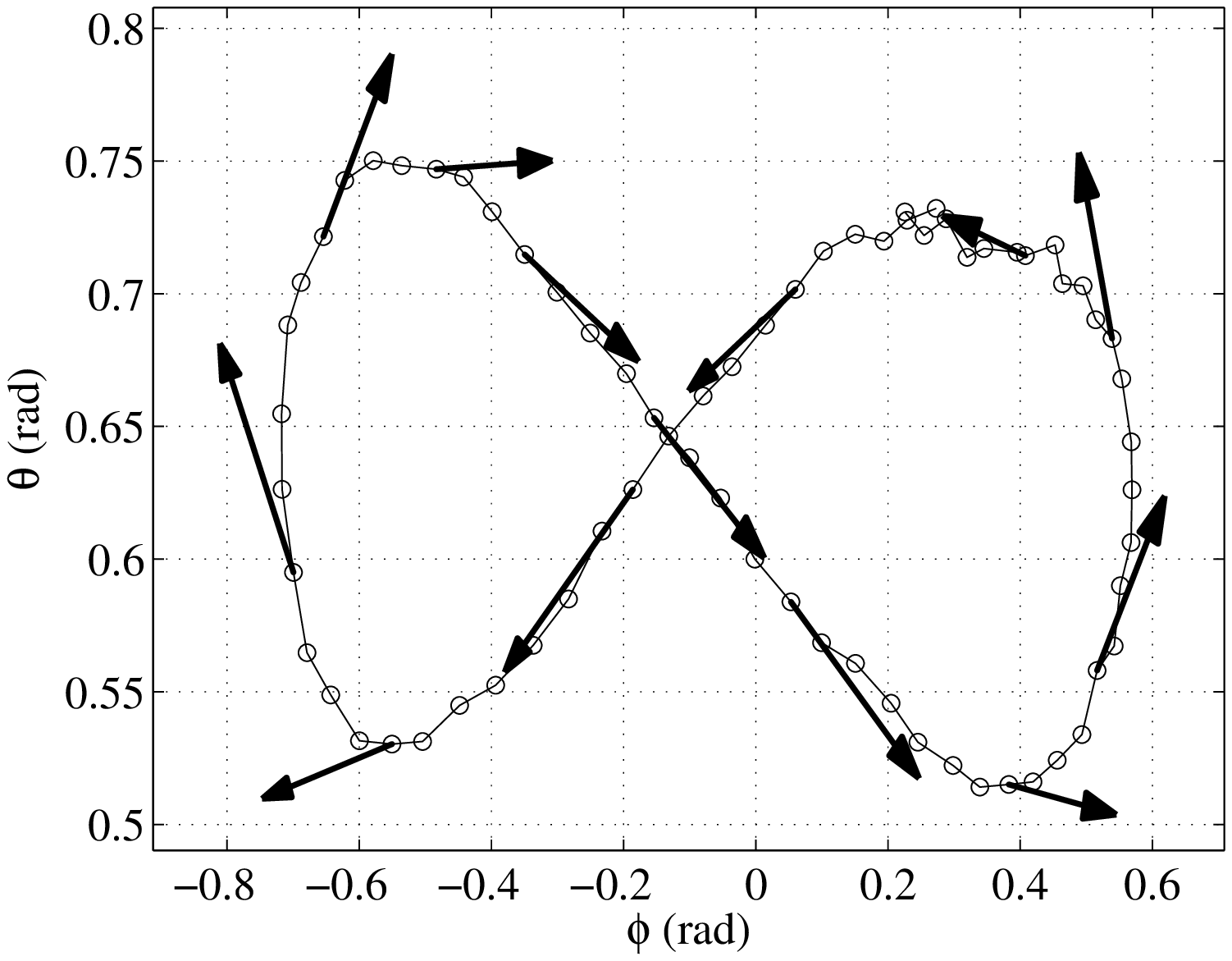}
\end{tabular}}
\caption{Experimental results. Examples of wing paths (solid lines with `$\circ$') and velocity angles (arrows) obtained with a (a) 6$\,$m$^2$, (b) 9$\,$m$^2$ and (c) 12$\,$m$^2$ wing and with about 2$\,$m/s (left) and 4$\,$m/s (right) wind speed measured at 4$\,$m above the ground.}
\label{F:trajectory_2}
\end{figure*}

We now focus on the results obtained with the third approach. Referring to section \ref{SS:tuning}, we choose the tuning parameters as $\lambda=500$ and $K_\gamma=[0.4,\,0.9]^T$. We first show an example of the time courses of the filtered velocity angle $\hat{\gamma}$ and its rate $\hat{\dot{\gamma}}$, see Fig. \ref{F:Filter_gamma}. It can be noted that the estimates are quite smooth, hence they are suitable to be used for feedback control, and that the derivative $\hat{\dot{\gamma}}$ has a rather small lag with respect to $\hat{\gamma}$.
\begin{figure}[hbt]
\centerline{
\includegraphics[bbllx=29mm,bblly=76mm,bburx=184mm,bbury=197mm,width=7.5cm,clip]{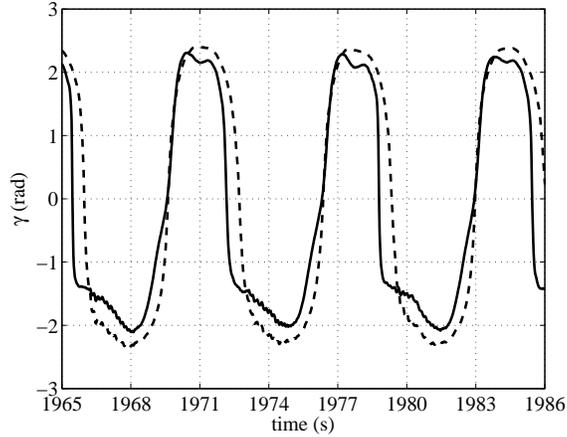}} \caption{Time courses of the filtered velocity angle $\hat{\gamma}$ obtained in the third approach with the IMU (solid line) and without (dashed line). Design parameters: $\lambda=500$, $K_\gamma=[0.4,\,0.9]^T$. Wing size: 12 m$^2$, wind speed: 2.3 m/s at 4$\,$m from the ground.}\label{F:gamma_w_wo_IMU}
\end{figure}
\begin{figure}[hbt]
\centerline{
\includegraphics[bbllx=26mm,bblly=75mm,bburx=185mm,bbury=200mm,width=7.5cm,clip]{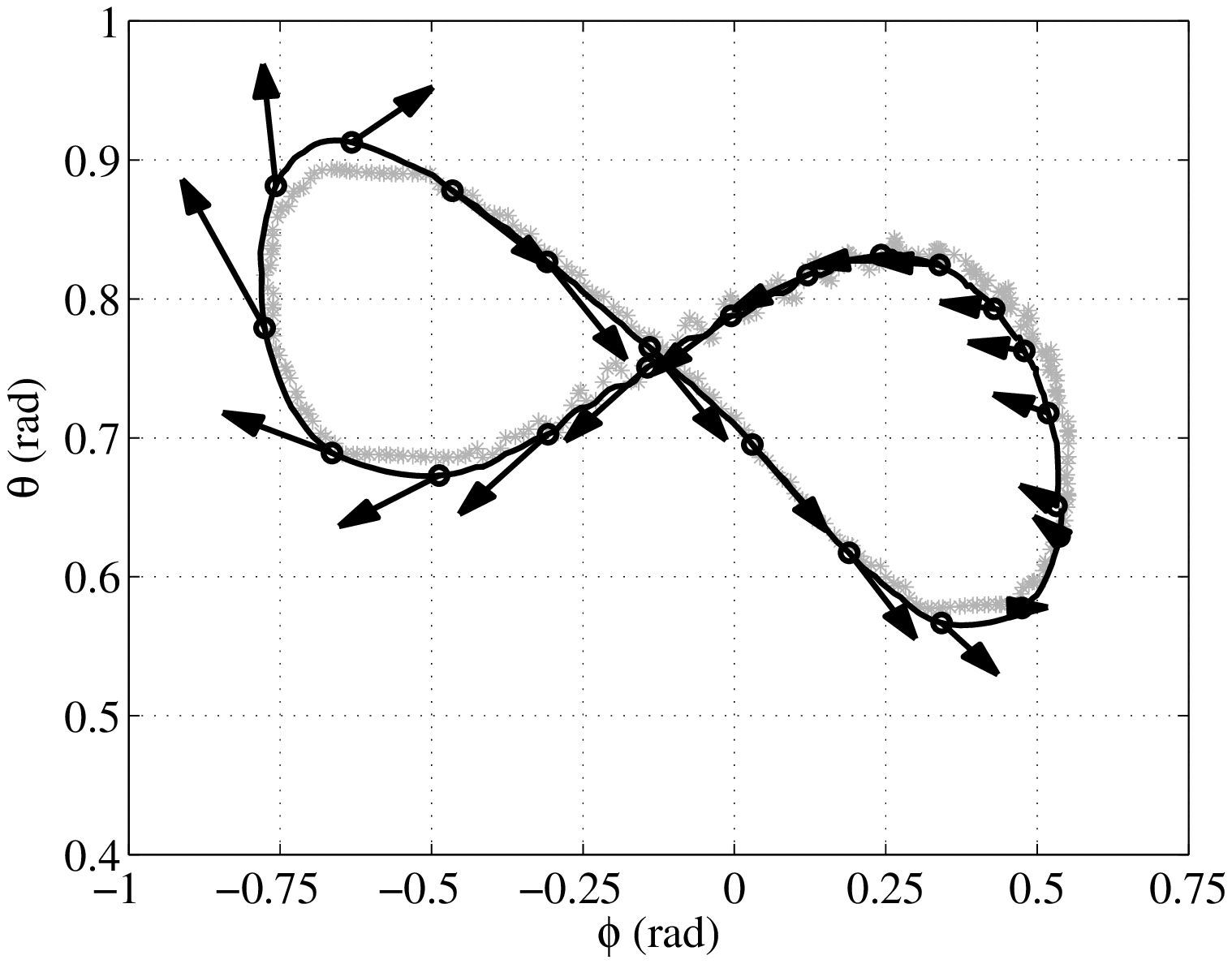}} \caption{Filtered position in the $(\phi,\theta)$ plane obtained with the third approach with the IMU (solid line and `$\,\circ\,$') and related velocity angle estimates (arrows). The raw position data is shown with gray `$\,*\,$'. Design parameters: $\lambda=500$, $K_\gamma=[0.4,\,0.9]^T$. Wing size: 12 m$^2$, wind speed: 2.3 m/s at 4$\,$m from the ground.}\label{F:loop_w_acc}
\end{figure}
\begin{figure}[hbt]
\centerline{
\includegraphics[bbllx=29mm,bblly=76mm,bburx=185mm,bbury=197mm,width=7.5cm,clip]{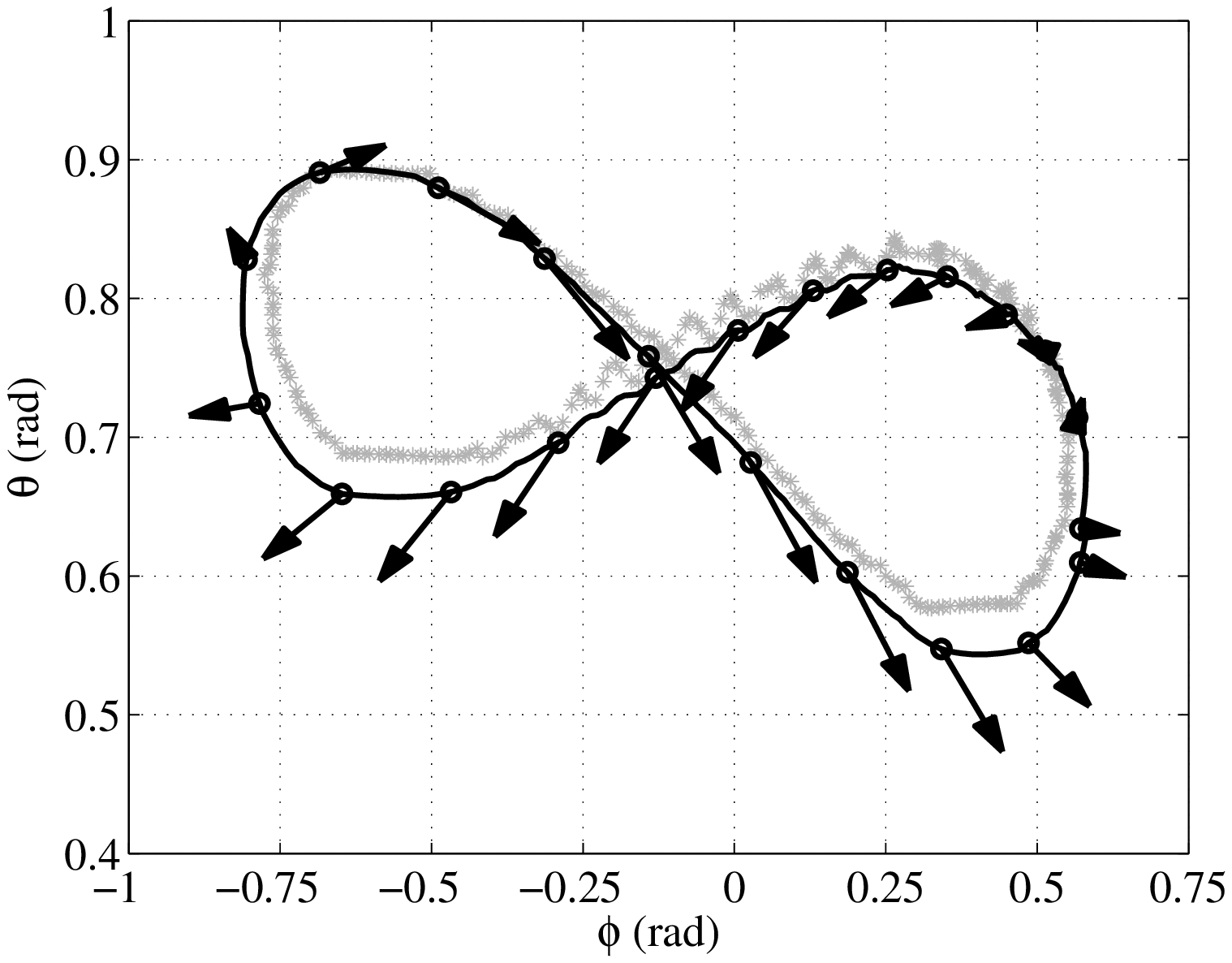}} \caption{Filtered position in the $(\phi,\theta)$ plane obtained with the third approach without the IMU (solid line and `$\,\circ\,$') and related velocity angle estimates (arrows). The raw position data is shown with gray `$\,*\,$'. Design parameters: $\lambda=500$, $K_\gamma=[0.4,\,0.9]^T$. Wing size: 12 m$^2$, wind speed: 2.3 m/s at 4$\,$m from the ground.}\label{F:loop_no_acc}
\end{figure}
In order to assess the contribution provided by the inertial sensors to the estimated quantities, we  show next, in Fig. \ref{F:gamma_w_wo_IMU}, the course of $\hat{\gamma}$ obtained either with the IMU (solid line) or without, i.e. considering $\vec{\hat{\ddot{p}}}_G=0$. It can be noted that in the second case, a quite significant lag is present in the estimate. Although the control system (designed according to \cite{FZMK13_arxiv}) can still operate satisfactorily in this case, its performance get worse. This conclusion is confirmed by the data reported in Figs.  \ref{F:loop_w_acc}-\ref{F:loop_no_acc}, which show the same flown path, together with its filtered position in the $(\phi,\,\theta)$ plane and the estimated velocity angles, again obtained either with or without the IMU. The raw position data given by the line angle sensor are reported in both figures as well. By comparing these two plots, it can be noted that the use of the IMU yields a better accuracy not only of the velocity angle estimate (compare the two figures e.g. for $\phi\simeq-0.75,\,\theta\simeq0.7$), but also of the filtered position, as shown by the larger errors between the raw data and the filtered ones in Figure \ref{F:loop_no_acc}. The reason of such a difference lies in the fact that the inertial sensors of the IMU allow to anticipate the wing's movement, hence reducing the lag in the filtered variables. In summary, while the third approach provides estimates that are good enough for feedback control also if the IMU is not used, the fusion of the inertial sensors of the IMU with the line angle measurement system yields the best performance.
In Fig. \ref{F:loop_w_acc}, it can be noted that the estimated velocity angle is quite accurate, with the gradient being almost always tangent to the estimated flying path. The length of the arrows in Fig. \ref{F:loop_w_acc}  is proportional to the wing's speed with respect to the ground. Similar good results have been obtained with all the three different wings, and in different wind conditions (see Fig. \ref{F:trajectory_2} for some examples), hence showing the robustness of the approach, deriving from the use of kinematic equations to design the filters. These features make the use of the proposed filtering technique and of $\hat{\gamma}^3$ most suited for feedback control, see \cite{Wing_movie} for a short movie clip concerned with autonomous flights carried out using such estimation algorithm.

\section{Discussion and conclusions}\label{S:conclusion}

We presented a study on the problem of sensor fusion for tethered wings to be used in airborne wind energy systems, focusing in particular on  crosswind flight conditions. We designed three algorithms to estimate the wing's position and velocity angle, using different sensors, and we applied them in experimental tests.

The approaches and the results presented in this paper are directly significant for small-scale airborne wind energy systems when operating with relatively short lines. Moreover, as already mentioned, the use of a kinematic model renders the approaches suitable for virtually any size and  type of wing. The main underlying assumption for the validity of the approaches is the presence of a straight tether linking the wing to the ground. In our experimental setup, with 30-m-long lines and force values ranging typically from 500$\,$N to 3000$\,$N in crosswind conditions, this was always the case. In such conditions, our results indicate that the techniques based on GPS are not usable in practice for the purpose of feedback control in crosswind motion. This holds in particular for the first approach for the position estimate, and for both the first and second approaches for the $\gamma$ estimate. Hence, our results suggest that the use of accurate and fast position measurement devices, like the line angle sensor, are essential to obtain high accuracy when the lines are relatively short w.r.t. the wing speed. More specifically, the GPS is affected by poor accuracy (which can be partially improved by exploiting the kinematic constraint given by the tether) and by a (time-varying) delay, typically about 0.2$\,$s but up to 1$\,$s in some cases. Finally, we also showed with experimental results that the fusion of inertial sensors with the position measurements yields the best performance in estimating the position, velocity and velocity angle of the wing, also when an accurate line angle sensor is used.

This situation is likely to change when longer lines are used, so that the additional line drag and weight might reduce the traction force exerted by the wing, giving rise to line sagging. Clearly, at which length of the lines this might happen depends on the wing size, efficiency and wind speed.  In such conditions, the accuracy of the line angle sensor decreases, hence making the third approach less effective, however on the other hand we expect the performance of the first and second approach to improve. In fact, it is reasonable to assume that the accuracy of the GPS would be the same in absolute values also with longer tether length, hence yielding smaller errors in terms of angular positions, which are the ones that matter for feedback control in several approaches \cite{ErSt12,FZMK13_arxiv}. This would hold particularly if the span of the flying path increases with the length of the lines (e.g. if the same trajectory in the $(\phi,\theta)$ plane is kept). Such a consideration suggests that the capability of tracking the wing with sufficiently good accuracy might be a design constraint for the size of the flown trajectories. Moreover, the use of GPS sensors with higher performance (e.g. differential GPS) can certainly improve the accuracy obtained with the first two approaches, which then could  be  mixed  with the third one. This represents an interesting line of research, that can be pursued only with a larger testing setup with full reel-out capabilities.

Focusing again on the third approach, eventual line sag could be also detected and corrected by exploiting the barometric altitude reading and the measure of the forces acting on the lines, as well as other additional measurements like onboard airspeed, which is also related to the generated force: from our own experience and from discussions with several researchers working in the field of airborne wind energy, line sag is almost absent during crosswind flight if the forces are large enough. This aspects provides further design and operation criteria for AWE systems: the chosen length of the tethers and reel-out speed should be matched with the characteristics of the wing (like size and efficiency) and with the wind conditions, in order to make sure that the load on the lines is large enough, relative to their length, to allow good position measurements from the ground with a line angle sensor. This point provides also a further reason to investigate the phenomenon of line sag, not only for the sake of deriving more accurate models for numerical simulations, but also for the purpose of estimation.

Last, we think that the use of larger and heavier wings would also yield better accuracy with the first two approaches, due to larger flight paths (whose radius increases approximately linearly with the wing span) and higher inertia that should lead to an  improvement in the relative GPS accuracy.

\end{document}